\documentstyle[12pt,amstex]{article}
\def\C{{\rm\kern.24em \vrule width.02em 
 height1.4ex depth-.05ex
\kern-.26em C}}
\def\R{{\rm I\kern-.20em R}}
\def\Z{{\rm\kern.26em \vrule width.02em height0.5ex depth0ex
 \kern.04em
 \vrule  width.02em height1.47ex depth-1ex \kern-.34em Z}}
\def\N{{\rm I\kern-.20em N}}
\def\Q{{\rm\kern.24em \vrule width.02em 
 height1.4ex depth-.05ex \kern-.26em Q}}

\def\beq{\begin{equation}}
\def\eeq{\end{equation}}
\def\beqa{\begin{eqnarray}}
\def\eeqa{\end{eqnarray}}

{\catcode `\| = 0 \catcode `\\ = 11 |long|gdef|filterouttext#1
\eskip{|endgroup}}

{\catcode `\| = 0 \catcode `\\ = 11 |gdef|bakslask{\}}
\def\eskip{\message{Warning: unmatched \bakslask eskip encountered}}
\begin{document}
 \begin{titlepage}
 .
 \vskip 3.5cm
 \begin{center}
  {\bf \Large Continuous and Discrete Homotopy Operators} \\
  {\bf \Large with Applications in Integrability Testing} \\
\vskip 1.3cm
       Willy Hereman, Michael Colagrosso, Ryan Sayers, Adam Ringler \\
          Department of Mathematical and Computer Sciences \\
                      Colorado School of Mines \\
                   Golden, CO 80401-1887, U.S.A. \\
\vskip 1cm
                     Bernard Deconinck, Michael Nivala \\
                   Department of Applied Mathematics \\
                        University of Washington \\
                      Seattle, WA 98195-2420, U.S.A. \\
\vskip 1cm
                            Mark S. Hickman \\
              Department of Mathematics and Statistics \\
             University of Canterbury, Private Bag 4800 \\
                      Christchurch, New Zealand. \\
\end{center}
\vskip 0.3cm
 \begin{center}
 {\sc In honor of Ryan Sayers (1982-2003)} 
 \end{center}
\vskip 1cm
 \begin{center}
 {\rm Research Support in Part by the National Science Foundation (NSF) under 
Grants Nos.\ DMS-9732069, DMS-9912293, CCR-9901929, FRG-DMS-0139093, and 
DMS-9810726.} \\
 \vskip 1cm
 \end{center}
\begin{center}
Manuscript prepared for Differential Equations with Symbolic Computation \\
Eds.: Dongming Wang and Zhiming Zheng
(2005). 
\end{center}
%
\begin{center}
2000 Mathematics Subject Classification \\
%
Primary: 37K05, 37J35, 35Q35; Secondary: 37K10, 35Q58, 37K60
%
\end{center}
\end{titlepage}
\vfill
\newpage
%
%
\section*{Abstract}

We introduce calculus-based formulas for the continuous Euler and  homotopy
operators. The 1D continuous homotopy operator automates integration by parts 
on the jet space.  Its 3D generalization allows one to invert the total
divergence operator. As a practical application, we show how the operators can
be used to symbolically compute local conservation laws of nonlinear systems
of  partial differential equations in multi-dimensions. 

By analogy to the continuous case, we also present concrete formulas  for the
discrete Euler and homotopy operators. Essentially, the discrete homotopy
operator carries out summation by parts. We use it to algorithmically invert
the forward difference operator. We apply the discrete operator to compute
fluxes of  differential-difference equations in (1+1) dimensions.

Our calculus-based approach allows for a straightforward implementation of the
operators in major computer algebra system, such as {\it Mathematica}  and {\it
Maple}. The symbolic algorithms for integration and summation by parts are 
illustrated with elementary examples. The algorithms to compute conservation
laws are illustrated with nonlinear  PDEs and their discretizations arising in
fluid dynamics and  mathematical physics.

%
%
\section{Introduction}
\label{introduction}
This chapter focuses on symbolic methods to compute polynomial conservation 
laws of partial differential equations (PDEs) in multi-dimensions and 
differential-difference equations (DDEs) (semi-discrete lattices).
For the latter we treat only (1+1) dimensional systems where time is 
continuous and the spacial variable has been discretized. 

There are several strategies to compute conservation laws of PDEs.
Some methods use a generating function \cite{MAandHSbook81}, 
which requires the knowledge of key pieces of the Inverse Scattering 
Transform \cite{MAandPCbook91}.
Other methods use Noether's theorem to get conservation laws from 
variational symmetries. 
More algorithmic methods, some of which circumvent the existence of a 
variational principle \cite{SAandGB02a,SAandGB02b,AKandFM00,TW02},
require the solution of a determining system of ODEs or PDEs.   
Despite their power, only a few of these methods have been implement 
in computer algebra systems (CAS), such as {\it Mathematica}, {\it Maple},
and {\it REDUCE}. 
See \cite{UGandWHjsc97,TW02} for reviews.

We advocate a more direct approach by building the candidate density 
as a linear combination (with constant coefficients) of terms that are 
uniform in rank (with respect to the scaling symmetry of the PDE). 
Although restricted to polynomial densities and fluxes, our method is 
entirely algorithmic and can be implemented in most CAS.
We refer the reader to \cite{UGandWHjsc97,UGandWHpd98} for details about 
an implementation in {\it Mathematica}, which can be downloaded from 
\cite{WHwebsite04}.
An implementation in {\it Maple} is also available \cite{BDwebsite04}.
%

Our earlier algorithm \cite{UGandWHjsc97,UGandWHpd98} worked only for 
nonlinear PDEs in one spacial variable.
In this chapter we present an algorithm that works for systems of PDEs 
in multi-dimensions that appear in fluid mechanics, elasticity, 
gas dynamics, general relativity, (magneto-)hydro-dynamics, etc.
%
%
The new algorithm produces densities in which all divergences and 
equivalent terms have been removed. 
An additional advantage of our methods to compute densities and fluxes 
is that they can be applied to nonlinear DDEs 
\cite{UGandWHpd98,WHetalcrm04,MHandWHprsa03}.
%
%
%

During the development of our methods we came across tools from the 
calculus of variations and differential geometry that deserve attention 
in their own right.
These tools are the variational derivative, the higher Euler operators, 
and the homotopy operator.

%
To set the stage, we address a few issues arising in multivariate calculus:
\newline
(i) To determine whether or not a vector field ${\bf F}$ is 
{\it conservative}, i.e.\ ${\mathbf F} = {\mbox {\boldmath $\nabla$}} f$ 
for some scalar field $f,$ one must verify that ${\mathbf F}$ is irrotational, 
that is ${\mbox {\boldmath $\nabla$}} \times {\mathbf F} = {\mathbf 0}.$
The field $f$ can be computed via standard integrations 
\cite[p.\ 518, 522]{JMandATbook88}.
\vskip 1pt
\noindent
(ii) To test if ${\mathbf F}$ is the curl of some vector field ${\mathbf G},$ 
one must check that ${\mathbf F}$ is {\it incompressible} or 
{\it divergence free}, 
i.e.\ ${\mbox {\boldmath $\nabla$}} \cdot {\mathbf F} = 0.$ 
The components of ${\mathbf G}$ result from solving a coupled system of 
first-order PDEs \cite[p.\ 526]{JMandATbook88}.
\vskip 2pt
\noindent
(iii) To verify whether or not a scalar field $f$ is the divergence of 
some vector function ${\mathbf F},$ no theorem from vector calculus
comes to the rescue.
Furthermore, the computation of ${\mathbf F}$ such that 
$f = {\mbox {\boldmath $\nabla$}} \cdot {\mathbf F}$ is a nontrivial matter.
In single variable calculus, it boils down to computing the primitive 
$F = \int f \, dx.$
In multivariate calculus, all scalar fields $f,$ including the components 
$F_i$ of vector fields ${\mathbf F} = (F_1,F_2,F_3),$ are functions of the 
independent variables $(x,y,z).$ 
In differential geometry one addresses the above issues in much greater 
generality. 
There, the functions $f$ and $F_i$ can depend on arbitrary functions 
$u(x,y,z),v(x,y,z),$ etc.\ and their mixed derivatives (up to a fixed order) 
with respect to the independent variables $(x,y,z).$ 
Such functions are called {\it differential functions} \cite{PObook93}.
As one might expect, carrying out the gradient-, curl-, or divergence-test 
requires advanced algebraic machinery. 
For example, to test whether or not 
$f = {\mbox {\boldmath $\nabla$}} \cdot {\mathbf F}$ requires the use of the  
variational derivative (Euler operator) in 3D. 
The actual computation of ${\mathbf F}$ requires integration by parts. 
That is where the homotopy operator and the variational complex come into play.

In 1D problems the continuous total homotopy 
operator\footnote{Hence forth, homotopy operator instead of total homotopy 
operator.}
reduces the problem of symbolic integration by parts to an integration 
with respect to a single variable. 
In 2D and 3D, the homotopy operator allows one to invert the total divergence 
operator and, again, reduce the problem to a single integration.
At the moment, no major CAS have reliable routines for integrating
expressions involving {\it unknown} functions and their derivatives.
%
%
%
%
As far as we know, no CAS offer a function to test if a differential 
function is a divergence.
Routines to symbolically invert the total divergence are certainly lacking.

The continuous homotopy operator is a universal, yet little known, 
tool that can be applied to many problems in which integration by parts 
(of arbitrary functions) in multi-variables plays a significant role.
We refer the reader to \cite[p.\ 374]{PObook93} for a history of the 
homotopy operator in the context of inverse problems of the calculus 
of variations.
%
%

A major motivation for writing this chapter is to demystify the 
homotopy operators. 
Therefore, we purposely avoid differential forms and abstract concepts 
such as the variational bicomplex.
Instead, we present down-to-earth calculus formulas for the homotopy 
operators which makes them readily implementable in major CAS.

By analogy with the continuous case, we also present formulas for the 
discrete versions of the Euler and homotopy operators.
The discrete homotopy operator is a powerful tool to invert the forward 
difference operator, whatever the application is.
It circumvents the necessary summation (by parts) by applying a set of 
variational derivatives followed by a one-dimensional integration with 
respect to an auxiliary parameter.
We use the homotopy operator to compute conserved fluxes of DDEs.
%
Numerous examples of such DDEs are given in \cite{YSbook03}.
Beyond DDEs, the discrete homotopy operator has proven to be useful in 
the study of difference equations \cite{PHandEMfcm04,EMandPHams02}.
To our knowledge, CAS offer no tools to invert the forward difference 
operator.
Once fully implemented, our discrete homotopy operator will overcome 
the shortcomings.

As shown in \cite{PHandEMfcm04,EMandPHams02}, the parallelism between
the continuous and discrete cases can be made rigorous and both theories 
can be formulated in terms of variational bicomplexes.
To make our work accessible to as wide an audience as possible, we do 
not explicitly use the abstract framework. 
Aficionados of {\it de Rham} complexes may consult 
\cite{IA92,IAbook04,IAandTDajm80,IKandAVbook98} and 
\cite{PHandEMfcm04,EMandPHams02,EMandRQ04}.
The latter papers cover the discrete variational bicomplexes.
\section{Examples of Nonlinear PDEs}
\label{continuousexamples}
We consider nonlinear systems of evolution equations in $(3+1)$ dimensions, 
\begin{equation}
\label{continuoussystem}
{\bf u}_t = {\bf G}({\bf u}, {\bf u}_x, {\bf u}_y, {\bf u}_z,
{\bf u}_{2x}, {\bf u}_{2y}, {\bf u}_{2z}, 
{\bf u}_{xy}, {\bf u}_{xz}, {\bf u}_{yz}, \dots),
\end{equation}
where ${\bf x} = (x,y,z)$ are space variables and $t$ is time.
The vector ${\bf u}(x,y,z,t)$ has $N$ components $u_i.$ 
In the examples we denote the components of ${\bf u}$ by $u,v,w,$ etc. 
Subscripts refer to partial derivatives. 
For brevity, we use ${\bf u}_{2x}$ instead of ${\bf u}_{xx},$ etc.\ and write 
${\bf G}({\bf u}^{(n)})$ to indicate that the differential function 
${\bf G}$ depends on derivatives up to order $n$ of ${\bf u}$ with 
respect to $x,y,$ and $z.$ 
For simplicity, we assume that ${\bf G}$ does not explicitly depend on 
${\bf x}$ and $t.$
No restrictions are imposed on the number of components, the order, and the 
degree of nonlinearity of the variables in ${\bf G}.$ 

We will predominantly work with polynomial systems, although systems 
involving one transcendental nonlinearity can also be handled.
If parameters are present in (\ref{continuoussystem}), they will be 
denoted by lower-case Greek letters. 
\vskip 2pt
\noindent
{\bf Example 1}:
The {\rm coupled Korteweg-de Vries (cKdV) equations} \cite{MAandPCbook91}, 
\begin{equation}
\label{ckdv}
u_t - 6 \beta u u_x + 6 v v_x - \beta u_{3x} = 0, 
\;\quad 
v_t + 3 u v_x + v_{3x} = 0, 
\end{equation}
where $\beta$ is a nonzero parameter, describes interactions of two waves 
with different dispersion relations. 
System (\ref{ckdv}) is known in the literature as the Hirota-Satsuma system.
It is completely integrable
\cite{MAandPCbook91,RHandJS1981} when $\beta=\frac{1}{2}.$
\vskip 2pt
\noindent
%
%
\vskip 4pt
\noindent
{\bf Example 2} 
The {\rm sine-Gordon (SG) equation} \cite{ABetalnc71,GLrmp71}, 
$u_{2t} - u_{2x} = \sin u,$ can be written as a system of evolution equations,
\begin{equation}
\label{sinegordon0}
u_t = v, \quad v_t = u_{2x} + \sin u. 
\end{equation}
This system occurs in numerous areas of mathematics and physics, 
ranging from surfaces with constant mean curvature to superconductivity.
%
\vskip 2pt
\noindent
{\bf Example 3}:
The (2+1)-dimensional {\rm shallow water wave (SWW) equations} \cite{PDpf03}, 
\begin{eqnarray}
\label{swwvector}
&& {\bf u}_t + ({\bf u} {\bf \cdot} {\mbox {\boldmath $\nabla$}} ) {\bf u} 
   + 2 {\mbox{\boldmath $\Omega$}} \times {\bf u} 
   = - {\mbox {\boldmath $\nabla$}} (h \theta ) 
     + {\textstyle \frac{1}{2}} h {\mbox {\boldmath $\nabla$}} \theta ,
\nonumber \\
&& \theta_t + {\bf u} {\bf \cdot} ({\mbox {\boldmath $\nabla$}} \theta) = 0, 
\quad
h_t + {\mbox {\boldmath $\nabla$}} {\bf \cdot} (h {\bf u}) = 0, 
\end{eqnarray}
describe waves in the ocean using layered models. 
Vectors ${\bf u} = u(x,y,t) {\bf i} + v(x,y,t) {\bf j}$ and 
${\mbox {\boldmath $\Omega$}} = 
\Omega {\bf k}$ are the fluid and angular velocities, respectively.
${\bf i}$, ${\bf j},$ and ${\bf k}$ are unit vectors along the $x$, $y,$ 
and $z$-axes. 
$\theta(x,y,t)$ is the horizontally varying potential temperature field,
and $h(x,y,t)$ is the layer depth.
The dot $(\cdot)$ stands for Euclidean inner product and 
${\mbox {\boldmath $\nabla$}} = 
\frac{\partial}{\partial x} {\bf i} + \frac{\partial}{\partial y} {\bf j}$
is the gradient operator.
System (\ref{swwvector}) is written in components as
\begin{eqnarray}
\label{sww}
&& u_t + u u_x + v u_y - 2 \Omega v + 
{\textstyle \frac{1}{2}} h \theta_x + \theta h_x \!=\! 0, \quad\quad
v_t + u v_x + v v_y + 2 \Omega u + 
{\textstyle \frac{1}{2}} h \theta_y + \theta h_y \!=\! 0, 
\nonumber \\
&& \theta_t + u \theta_x + v \theta_y \!=\! 0, \quad\quad
h_t + h u_x + u h_x + h v_y + v h_y \!=\! 0.
\end{eqnarray}
%
%
\section{Key Definitions--Continuous Case}
\label{definitionsPDEs}
%
{\bf Definition}: 
System (\ref{continuoussystem}) is said to be {\it dilation invariant} 
if it is invariant under a scaling (dilation) symmetry.  
%
\vskip 2pt
\noindent
{\bf Example}: 
The cKdV system (\ref{ckdv}) is invariant under the scaling symmetry
\begin{equation}
\label{ckdvscale}
(x, t, u, v) \rightarrow 
(\lambda^{-1} x, \lambda^{-3} t, \lambda^2 u, \lambda^2 v), 
\end{equation}
where $\lambda$ is an arbitrary scaling parameter.
%
%
\vskip 2pt
\noindent
{\bf Definition}: 
We define the {\it weight}, $W$, of a variable as the exponent 
of $\lambda$ that multiplies the variable.
\vskip 2pt
\noindent
{\bf Example}:
We will always replace $x$ by $\lambda^{-1} x.$  
Thus, $W(x) = -1$ or $W({\partial/\partial x}) = 1.$ 
From (\ref{ckdvscale}), we have $W({\partial/\partial t}) = 3$ and 
$W(u) = W(v) = 2$ for the cKdV equations.
%
\vskip 2pt
\noindent
{\bf Definition}: 
The {\it rank} of a monomial is defined as the total weight of the monomial.
An expression (or equation) is {\it uniform in rank} if its monomial 
terms have equal rank.
%
\vskip 2pt
\noindent
{\bf Example}:
Coincidentally, all monomials in {\it both} equations of (\ref{ckdv}) 
have rank 5. 
Thus, (\ref{ckdv}) is uniform in rank.
The ranks of the equations in (\ref{continuoussystem}) may differ from 
each other. 
\vskip 2pt
\noindent
%
Conversely, requiring {\rm uniformity in rank} for each equation in 
(\ref{continuoussystem}) allows one to compute the weights of the  
variables (and thus the scaling symmetry) with linear algebra. 
\vskip 2pt
\noindent
{\bf Example}:
For the cKdV equations (\ref{ckdv}), one has
\begin{eqnarray}
\label{ckdvweightequations}
W(u) + W({\partial/\partial t}) 
\!&\!=\!&\! 2 W(u) + 1 = 2 W(v) + 1 = W(u) + 3, 
\nonumber \\
W(v) + W({\partial/\partial t}) 
\!&\!=\!&\! W(u) + W(v) + 1 = W(v) + 3,
\end{eqnarray}
which yields 
$ W(u) = W(v) = 2,\, W({\partial/\partial t}) = 3, $
which leads to (\ref{ckdvscale}).
%
\vskip 2pt
\noindent
Dilation symmetries, which are special Lie-point symmetries, are common to
many nonlinear PDEs. 
However, non-uniform PDEs can be made uniform by extending the set of 
dependent variables with auxiliary parameters with appropriate weights. 
Upon completion of the computations one can set these parameters to one.
\vskip 2pt
\noindent
{\bf Example}:
The sine-Gordon equation (\ref{sinegordon0}) is not uniform in rank unless 
we replace it by 
\begin{equation}
\label{sinegordon}
u_t = v, \quad v_t = u_{2x} + \alpha \sin u, \quad\; \alpha \in \R.
\end{equation}
Using the Maclaurin series for the $\sin$ function, 
uniformity in rank requires 
\begin{eqnarray}
\label{sinegordonweightequations}
\!\!W(u) + W({\partial/\partial t}) \!\!&\!\!=\!\!&\!\! W(v), 
\nonumber \\
\!\!\!W(v) \!+\! W({\partial/\partial t}) \!\!&\!\!=\!\!&\!\! W(u) \!+\! 2 
\!=\!W(\alpha) \!+\! W(u) \!=\! 
W(\alpha) \!+\! 3 W(u) \!=\! W(\alpha) \!+\! 5 W(u) \!=\! \cdots . 
\end{eqnarray}
This forces us to set $W(u) = 0.$ Then, $W(\alpha) = 2.$
By allowing the parameter $\alpha$ to scale, (\ref{sinegordon}) becomes 
scaling invariant under the symmetry
\begin{equation}
\label{sinegordoncale}
(x, t, u, v, \alpha) \rightarrow 
(\lambda^{-1} x, \lambda^{-1} t, \lambda^0 u, 
\lambda^1 v, \lambda^2 \alpha), 
\end{equation}
corresponding to 
$W({\partial/\partial x}) = W({\partial/\partial t}) = 1, 
W(u) = 0, W(v) = 1, W(\alpha) = 2.$
The first and second equations in (\ref{sinegordon}) are uniform of 
ranks 1 and 2, respectively.
\vskip 2pt
\noindent
{\bf Definition}: 
System (\ref{continuoussystem}) is called {\it multi-uniform} in rank 
if it admits more than one dilation symmetry (which is not the result of 
adding auxiliary parameters with weights).
%
\vskip 2pt
\noindent
{\bf Example}:
Uniformity in rank for the SWW equations (\ref{sww}) requires, after some 
algebra, that
%
%
\begin{eqnarray}
\label{swwweightequations}
&& W({\partial/\partial t}) =
W(\Omega), \quad 
W({\partial/\partial y}) = W({\partial/\partial x}) = 1, 
\quad W(u) = W(v) = W(\Omega) - 1, 
\nonumber \\
&& W(\theta) =
2 W(\Omega) - W(h) - 2, 
\end{eqnarray} 
where $W(h)$ and $W(\Omega)$ remain free. 
The SWW system is thus multi-uniform. 
The symmetry 
\begin{equation}
\label{swwscale}
(x, y, t, u, v, \theta, h, \Omega) \rightarrow 
(\lambda^{-1} x, \lambda^{-1} y, \lambda^{-2} t, 
\lambda u, \lambda v, \lambda \theta, \lambda h, \lambda^2 \Omega),
\end{equation}
which is most useful for our computations later on, corresponds to 
$W({\partial/\partial x}) \!=\! W({\partial/\partial y}) \!=\!1, 
W({\partial/\partial t}) \!=\!2, W(u) \!=\! W(v) \!=\!1, 
W(\theta) \!=\!1, W(h) \!=\!1,$ and $W(\Omega) \!=\!2.$
A second symmetry,
\begin{equation}
\label{swwscale0}
(x, y, t, u, v, \theta, h, \Omega) \rightarrow 
(\lambda^{-1} x, \lambda^{-1} y, \lambda^{-2} t, 
\lambda u, \lambda v, \lambda^2 \theta, \lambda^0 h, \lambda^2 \Omega), 
\end{equation}
matches
$W({\partial/\partial x})\!\!=\!\!W({\partial/\partial y})\!\!=\!\!1, 
W({\partial/\partial t})\!\!=\!\!2, W(u)\!\!=\!\!W(v)\!\!=\!\!1, 
W(\theta)\!\!=\!\!2, W(h)\!\!=\!\!0, W(\Omega)\!\!=\!\!2.$
%
%
\vspace{-0.50cm}
\section{Conserved Densities and Fluxes of Nonlinear PDEs}
\label{densfluxPDEs}
%
{\bf Definition}: 
A scalar differential function $\rho({\bf u}^{(n)})$ is a conserved 
{\it density} if there exists a vector differential function 
${\bf J}({\bf u}^{(m)}),$ called the associated {\it flux}, such that
%
%
\begin{equation} 
\label{pdeconslaw} 
{\rm D}_{t} \, \rho + {\rm Div} \, {\bf J} = 0
\end{equation}
is satisfied on the solutions of (\ref{continuoussystem}). 
Equation (\ref{pdeconslaw}) is called a 
local\footnote{We only compute densities and fluxes free of integral terms.}
{\it conservation law} \cite{PObook93}, 
and {\rm Div} is called the total divergence
\footnote{Gradient, curl, and divergence are in rectangular coordinates.}. 
Clearly, 
$ {\rm Div} \, {\bf J} 
= ( {\rm D}_x, {\rm D}_y, {\rm D}_z ) \cdot ( J_1, J_2, J_3 ) 
= {\rm D}_x J_1 + {\rm D}_y J_2 + {\rm D}_z J_3. $ 
%
In the case of one spacial variable $(x),$ (\ref{pdeconslaw}) reduces to 
\begin{equation}
\label{1dpdeconslaw}
{\rm D}_t \, \rho + {\rm D}_x J = 0,
\end{equation}
where both density $\rho$ and flux $J$ are scalar differential functions. 
In the 1D case, 
\begin{equation}
\label{operatordt}
{\rm D}_t \rho(u^{(n)}) = \frac{\partial \rho}{\partial t} 
+ \sum_{k=0}^{n} \frac{\partial \rho}{\partial u_{kx}} {\rm D}_x^k u_t.
\end{equation}
where $u^{(n)}$ is the highest order term present in $\rho.$
Upon replacement of $u_t, u_{tx},$ etc.\ from $u_t = G,$ one gets
\begin{equation}
\label{operatordtfrechet}
{\rm D}_t \rho = 
\frac{\partial \rho}{\partial t} + \rho(u)^{\prime}[G],
\end{equation}
where $\rho(u)^{\prime}[G]$ is the Fr\'echet derivative of $\rho$ 
in the direction of $G.$
Similarly, 
\begin{equation}
\label{operatordx}
{\rm D}_x J(u^{(m)}) 
= \frac{\partial J}{\partial x} 
+ \sum_{k=0}^{m} \frac{\partial J}{\partial u_{kx}} u_{(k+1)x}.
\end{equation}
%
The generalization of (\ref{operatordt}) and (\ref{operatordx}) to 
multiple dependent variable is straightforward.
For example, taking ${\bf u} = (u,v),$
\begin{eqnarray}
\label{operatordtforuandv}
{\rm D}_t \rho(u^{(n_1)},v^{(n_2)}) 
&=& \frac{\partial \rho}{\partial t} 
+ \sum_{k=0}^{n_1} \frac{\partial \rho}{\partial u_{kx}} {\rm D}_x^k u_t 
+ \sum_{k=0}^{n_2} \frac{\partial \rho}{\partial v_{kx}} {\rm D}_x^k v_t, 
\\
\label{operatordxforuandv}
\label{operatordxuandv}
{\rm D}_x J(u^{(m_1)},v^{(m_2)}) 
&=& \frac{\partial J}{\partial x} 
+ \sum_{k=0}^{m_1} \frac{\partial J}{\partial u_{kx}} u_{(k+1)x} 
+ \sum_{k=0}^{m_2} \frac{\partial J}{\partial v_{kx}} v_{(k+1)x}.
\end{eqnarray}
We will ignore densities and fluxes that explicitly depend on ${\bf x}$ 
and $t.$
If ${\bf G}$ is polynomial then most, but not all, densities and fluxes 
are also polynomial.
%
%
\vskip 2pt
\noindent
{\bf Example}:
The first four density-flux pairs for the cKdV equations (\ref{ckdv}) are
\begin{eqnarray}
\label{ckdvconslaw1}
\rho^{(1)} \!&\!=\!&\! u, \quad \quad \quad \quad\;\;
J^{(1)} = -3 \beta u^2 + 3 v^2 - \beta u_{2x}, 
\quad\quad\quad\quad\quad\quad\quad\quad\; ({\rm any} \; \beta) 
\\
\label{ckdvconslaw2}
\rho^{(2)} \!&\!=\!&\! u^2 -2 v^2, 
\quad\;\,
J^{(2)} =
- 4 \beta u^3 + \beta u_x^2 - 2 \beta u u_{2x} + 2 v_x^2 - 4 v v_{2x}, 
\quad\! ({\rm any} \, \beta )
\\
\label{ckdvconslaw3}
\rho^{(3)} \!&\!=\!&\! u v,
\quad\quad\quad\quad
J^{(3)} = 3 u^2 v + 2 u^3 - u_x v_x + u_{2x} v + u v_{2x}, 
\quad\quad\; (\beta = -1) 
\end{eqnarray}
and 
\begin{eqnarray}
\label{ckdvconslaw4}
\rho^{(4)} \!&\!=\!&\! 
(1 + \beta ) u^3 - 3 u v^2 
- \textstyle{\frac{1}{2}}(1+\beta) {u_x}^2 
+ 3 {v_x}^2, 
\\
\label{ckdvconslaw4j}
J^{(4)} \!&\!=\!&\! 
-\textstyle{\frac{9}{2}} \beta (1 + \beta ) u^4 
+ 9 \beta u^2 v^2 
- \textstyle{\frac{9}{2}} v^4
+ 6 \beta (1 + \beta ) u u_x^2 
- 3 \beta (1 + \beta ) u^2 u_{2x}
\nonumber \\
&& + 3 \beta v^2 u_{2x} 
- \textstyle{\frac{1}{2}} \beta (1 + \beta ) u_{2x}^2 
+ \beta (1 + \beta ) u_x u_{3x} 
- 6 \beta v u_x v_x + 12 u v_x^2 
\nonumber \\
&& - 6 u v v_{2x} 
- 3 v_{2x}^2 
+ 6 v_x v_{3x} 
\quad\quad\quad\quad\quad\quad\quad\quad\quad\;\; (\beta \ne -1).  
\end{eqnarray}
%
The above densities are uniform in ranks $2,4$ and $6.$
Both $\rho^{(2)}$ and $\rho^{(3)}$ are of rank 4. 
The corresponding fluxes are also uniform in rank with ranks $4,6,$ and $8.$
In \cite{UGandWHjsc97}, we listed a few densities of rank $\ge 8,$ 
which only exist when $\beta = \frac{1}{2}.$

In general, if in (\ref{1dpdeconslaw}) ${\rm rank}\, \rho=R$ 
then ${\rm rank}\, J = R + W({\partial/\partial t}) - 1 .$ 
All the terms in (\ref{pdeconslaw}) are also uniform in rank.
This comes as no surprise since the conservation law (\ref{pdeconslaw})
holds on solutions of (\ref{continuoussystem}), hence it ``inherits"
the dilation symmetry of (\ref{continuoussystem}).
\vskip 2pt
\noindent
{\bf Example}:
The first few densities \cite{PAthesis03,RDandRBprsa77} for the 
sine-Gordon equation (\ref{sinegordon}) are
\begin{eqnarray}
\label{sinegordonconslaw1}
\rho^{(1)} \!&=&\! 
2 \alpha \cos u + v^2 + {u_x}^2,
\quad\quad\quad\quad\quad
J^{(1)} = - 2 u_x v,  
\\
\label{sinegordonconslaw2}
\rho^{(2)} \!&=&\! 2 u_x v,
\quad\quad\quad\quad\quad\quad\quad\quad\quad\quad\quad\;
J^{(2)} = 2 \alpha \cos u - v^2 - {u_x}^2, 
\\
\label{sinegordonconslaw3}
\rho^{(3)} \!&=&\! 
6 \alpha v u_x \cos u + v^3 u_x + v {u_x}^3 - 8 v_x u_{2x}, 
\\
\label{sinegordonconslaw4}
\rho^{(4)} \!&=&\! 
2 \alpha^2 \cos^2 u  - 2 \alpha^2 \sin^2 u + 
4 \alpha v^2 \cos u + 20 \alpha {u_x}^2 \cos u 
+ v^4 + 6 v^2 {u_x}^2 + {u_x}^4
\nonumber \\ 
\!\!
&& - 16 {v_x}^2 - 16 u_{2x}^2. 
\end{eqnarray}
$J^{(3)}$ and $J^{(4)}$ are not shown due to length.
Again, all densities and fluxes are uniform in rank 
(before $\alpha$ is set equal to 1).
\vskip 2pt
\noindent
{\bf Example}: 
The first few conserved densities and fluxes for the SWW equations 
(\ref{sww}) are
\begin{eqnarray}
\label{swwconslaw}
\rho^{(1)} \!\!&\!\!=\!\!&\!\! h, 
\;\;
{\bf J}^{(1)} \!=\!
\left(
\begin{array}{c}
 u h  \\
 v h  
\end{array}
\right), 
\;\;
\rho^{(2)} \!=\! h \theta , 
\;\;
{\bf J}^{(2)} \!=\!
\left(\!
\begin{array}{c}
 u h \theta \\
 v h \theta 
\end{array}
\!\right), 
\;\;
\rho^{(3)} \!=\! h \theta^2 , 
\;\;
{\bf J}^{(3)} \!=\!
\left(\!
\begin{array}{c}
 u h \theta^2  \\
 v h \theta^2 
 \end{array}
 \!\right),
\\
\label{swwconslawbis}
\rho^{(4)} \!\!&\!\!=\!\!&\!\! (u^2 + v^2) h + h^2 \theta , 
\;\;
{\bf J}^{(4)} \!=\!
\left(\!
\begin{array}{c}
 u^3 h + u v^2 h + 2 u h^2 \theta \\
 v^3 h + u^2 v h + 2 v h^2 \theta 
\end{array}
\!\right),
\;\;
\rho^{(5)} \!=\!
v_x \theta - u_y \theta + 2 \Omega \theta,
\\
\label{swwconslawtris}
{\bf J}^{(5)} 
\!\!&\!\!=\!\!&\!\! 
\frac{1}{6}\!\left(\!
\begin{array}{c}
12 \Omega u \theta - 4 u u_y \theta + 6 u v_x \theta + 2 v v_y \theta 
+ u^2 \theta_y + v^2 \theta_y - h \theta \theta_y + h_y \theta^2 
\\
12 \Omega v \theta + 4 v v_x \theta - 6 v u_y \theta - 2 u u_x \theta 
- u^2 \theta_x - v^2 \theta_x + h \theta \theta_x - h_x \theta^2 
\end{array}
\!\right). 
\end{eqnarray}
All densities and fluxes are multi-uniform in rank, which will substantially 
simplify the computation of the densities.
Under either of the two scales, (\ref{swwscale}) or (\ref{swwscale0}), 
${\rm rank}({\bf J}) = {\rm rank}(\rho) + 1.$
With the exception of $\rho^{(2)}$ and ${\bf J}^{(2)},$ the ranks of the
densities under scales (\ref{swwscale}) and (\ref{swwscale0}) differ by one.
The same holds for the fluxes. 
%
%
\section{Tools from the Calculus of Variations}
\label{toolsPDEs}
In this section we introduce the variational derivative (Euler operator),
the higher Euler operators from the calculus of variations, and the homotopy 
operator from homological algebra. 
These tools will be applied to the computation of densities and fluxes in 
Section~\ref{applicationsPDEs}.
%
\subsection{Continuous Variational Derivative (Euler Operator)}
\label{zeroeulercont}
{\bf Definition}: 
A scalar differential function $f$ is a {\it divergence} if and only if 
there exists a vector differential function ${\bf F}$ such that 
$f = {\rm Div} \, {\bf F}.$ 
In 1D, we say that a differential function $f$ is 
{\it exact}\footnote{We do not use {\it integrable} to avoid confusion with
complete integrability from soliton theory.} 
if and only if there exists a scalar differential function $F$ such that 
$f = {\rm D}_x F.$ 
Obviously, $F = {\rm D}_x^{-1}(f) = \int f \, dx$ is then the primitive 
(or integral) of $f.$
\vskip 2pt
\noindent
{\bf Example}: 
Consider
\begin{equation}
\label{fcontinuous}
f = 3 \, u_x \, v^2 \, \sin u - u_x^3 \, \sin u
- 6 \, v \, v_x \, \cos u + 2 \, u_x \, u_{2x} \, \cos u
+ 8 \, v_x v_{2x},
\end{equation}
which we encountered \cite{PAthesis03} while computing conservation laws 
for (\ref{sinegordon}).
$f$ is exact. 
Indeed, upon integration by parts (by hand), one gets
\begin{equation}
F = 4 \, v_x^2 + u_x^2 \, \cos u -3 \, v^2 \, \cos u.
\end{equation}
Most CAS, including {\it Mathematica}, 
{\it Maple}\footnote{Version 9.5 of {\it Maple} can integrate such
expressions as a result of our interactions with the developers.} 
and {\it Reduce}, fail this elementary integration!
\vskip 2pt
\noindent
{\bf Example}: 
Consider 
\begin{equation}
\label{divergenceuv}
f = u_x v_y - u_{2x} v_y - u_y v_x + u_{xy} v_x.
\end{equation}
It is easy to verify that $f = {\rm Div} \, {\bf F}$ with 
\begin{equation}
\label{vectorfordivergenceuv}
{\bf F} = (u v_y - u_x v_y, -u v_x + u_x v_x).
\end{equation}
As far as we know, the leading CAS have no tools to compute ${\bf F}.$
Three questions arise: 
\vskip 2pt
\noindent
(i) Under what conditions for $f$ does a closed form for ${\bf F}$ exist?
\vskip 1pt
\noindent
(ii) If $f$ is a divergence, what is it the divergence of?
\vskip 1pt
\noindent
(iii) Avoiding integration by parts, how can one design a fast algorithm to 
compute ${\bf F}?$
\vskip 2pt
\noindent
To answer these questions we use the following tools from the calculus of 
variations: the variational derivative (Euler operator), 
its generalizations (higher Euler operators), and the homotopy operator.
%
%
%
%
\vskip 2pt
\noindent
{\bf Definition}:
The {\it variational derivative} (Euler operator), 
${\cal L}^{({\bf 0})}_{ {\bf u}({\bf x})}$, is defined 
\cite[p.\ 246]{PObook93} by 
\begin{equation}
\label{zeroeulervectoruvectorx}
{\cal L}^{({\bf 0})}_{{\bf u}({\bf x})} = 
\sum_J (-{\rm D})_J \frac{\partial}{\partial {\bf u}_J}, 
\end{equation}
where the sum is over all the unordered multi-indices $J$ 
\cite[p.\ 95]{PObook93}.
For example, in the 2D case the multi-indices corresponding to 
second-order derivatives can be identified with 
$\{2x,2y,2z,xy,xz,yz\}.$
Obviously, $(-D)_{2x} = {\rm D}_x^2, \; (-D)_{xy} = {\rm D}_x {\rm D}_y,$ 
etc.\ For notational details see \cite[p.\ 95, p.\ 108, p.\ 246]{PObook93}.
\vskip 3pt
\noindent
With applications in mind, we give explicit formulas for the variational 
derivatives in 1D, 2D, and 3D. 
For scalar component $u$ they are
\begin{equation}
\label{zeroeulerscalarux}
\!\!\!\!\!\!\!\!\!\! {\cal L}^{(0)}_{u(x)} 
=
\sum_{k=0}^{\infty} (-{\rm D}_x)^k \frac{\partial}{\partial u_{kx} } 
=
\frac{\partial}{\partial {u} } 
- {\rm D}_x \frac{\partial}{\partial {u_x} }
+ {\rm D}_{x}^2 \frac{\partial }{\partial {u_{2x}} } 
- {\rm D}_{x}^3 \frac{\partial }{\partial {u_{3x}} } 
+ \cdots , 
\end{equation}
%
\begin{eqnarray}
\label{zeroeulerscalaruxy}
\!\!\!\!\!\!\!\!\!\! {\cal L}^{(0,0)}_{u(x,y)} 
\!&\!\!=\!\!&\! 
\sum_{k_{x}=0}^{\infty} \; \sum_{k_{y}=0}^{\infty}
(-{\rm D}_x)^{k_{x}} (-{\rm D}_y)^{k_{y}} 
\frac{\partial }{\partial u_{k_{x}x \, k_{y}y}}
\nonumber \\
\!&\!\!=\!\!&\! 
\frac{\partial}{\partial {u} } 
-\! {\rm D}_x \frac{\partial}{\partial {u_x} }
-\! {\rm D}_y \frac{\partial}{\partial {u_y} }
+\! {\rm D}_{x}^2 \frac{\partial }{\partial {u_{2x}} } 
+\! {\rm D}_{x} {\rm D}_{y} \frac{\partial }{\partial {u_{xy}} } 
+\! {\rm D}_{y}^2 \frac{\partial }{\partial {u_{2y}} } 
-\! {\rm D}_{x}^3 \frac{\partial }{\partial{ u_{3x}} } 
-\! \cdots ,
\end{eqnarray}
%
%
%
and
\begin{eqnarray}
\label{zeroeulerscalaruxyz}
\!\!\!\!\!\!\!\!\!\! {\cal L}^{(0,0,0)}_{u(x,y,z)} 
\!&\!\!=\!\!&\! 
\sum_{k_{x}=0}^{\infty} \; 
\sum_{k_{y}=0}^{\infty} \;
\sum_{k_{z}=0}^{\infty} \,
(-{\rm D}_x)^{k_{x}} \!
(-{\rm D}_y)^{k_{y}} \!
(-{\rm D}_z)^{k_{z}} \!
\frac{\partial }{\partial {\bf u}_{k_{x}x \, k_{y}y \, k_{z}z}}
\nonumber \\
\!&\!\!=\!\!&\! 
\frac{\partial}{\partial{u}} 
- {\rm D}_x \frac{\partial}{\partial {u_x} }
- {\rm D}_y \frac{\partial}{\partial {u_y} }
- {\rm D}_z \frac{\partial}{\partial {u_z} }
+ {\rm D}_{x}^2 \frac{\partial }{\partial {u_{2x}} } 
+ {\rm D}_{y}^2 \frac{\partial }{\partial {u_{2y}} } 
+ {\rm D}_{z}^2 \frac{\partial }{\partial {u_{2z}} } 
\nonumber \\
&& + \, {\rm D}_{x} {\rm D}_{y} \frac{\partial }{\partial {u_{xy}} } 
+ {\rm D}_{x} {\rm D}_{z} \frac{\partial }{\partial {u_{xz}} } 
+ {\rm D}_{y} {\rm D}_{z} \frac{\partial }{\partial {u_{yz}} } 
- {\rm D}_{x}^3 \frac{\partial }{\partial {u_{3x}} } 
- \cdots .
\end{eqnarray}
Note that ${\bf u}_{k_{x}x \, k_{y}y}$ stands for 
${\bf u}_{xx \cdots x \, yy \cdots y}$ where $x$ is repeated $k_x$ times 
and $y$ is repeated $k_y$ times. 
Similar formulas hold for components $v,w,$ etc.
\vskip 3pt
\noindent
The first question is then answered by the following theorem 
\cite[p.\ 248]{PObook93}.
\vskip 2pt
\noindent
{\bf Theorem}:
A necessary and sufficient condition for a function $f$ 
to be a divergence, i.e.\ there exists a ${\bf F}$ so that 
$f = {\rm Div} \, {\bf F},$ is that 
${\cal L}_{{\bf u}({\bf x})}^{(0)} (f) \equiv 0.$
In other words, the Euler operator annihilates divergences, just as 
the divergence annihilates curls, and the curl annihilates gradients.
\vskip 1pt
\noindent
If, for example, ${\bf u} = (u,v)$ then both ${\cal L}_{u({\bf x})}^{(0)} (f)$ 
and ${\cal L}_{v({\bf x})}^{(0)} (f)$ must vanish identically.
For the 1D case, the theorem says that a differential function $f$ 
is exact, i.e.\ there exists a $F$ so that $f = D_x F,$ 
if and only if ${\cal L}^{(0)}_{{\bf u}(x)} (f) \equiv 0.$
%
\vskip 4pt
\noindent
{\bf Example}:
To test the exactness of $f$ in (\ref{fcontinuous})
%
%
which involves just one independent variable $x,$ we apply the zeroth 
Euler operator (\ref{zeroeulervectoruvectorx}) to $f$ for each component 
of ${\bf u} = (u,v)$ separately.
For component $u$ (of order 2), one computes 
\begin{eqnarray}
{\cal L}^{(0)}_{u(x)} (f) 
\!&\!=\!&\!
\frac{\partial f}{\partial u} - {\rm D}_x \frac{\partial f}{\partial u_x}
+ {\rm D}_{x}^2 \frac{\partial f}{\partial u_{2x}}
\nonumber \\
\!&\!=\!&\! 
 3  u_x \, v^2 \, \cos u - u_x^3 \, \cos u + 6  v \, v_x \, \sin u
 - 2  u_x \, u_{2x} \, \sin u 
\nonumber \\
&& - {\rm D}_x [  3  v^2 \, \sin u - 3  u_x^2 \, \sin u + 2 u_{2x} \, \cos u ]
 + {\rm D}_x^2 [  2  u_x \, \cos u ] 
\nonumber \\
\!&\!=\!&\! 
 3  u_x \, v^2 \, \cos u - u_x^3 \, \cos u + 6 v \, v_x \, \sin u
 - 2 u_x \, u_{2x} \, \sin u 
\nonumber \\
&& 
- [ 3 u_x v^2 \, \cos u + 6 v \, v_x \, \sin u - 3 u_x^3 \, \cos u 
    - 6  u \, u_{2x} \, \sin u 
\nonumber \\
&&  - 2 u_x \, u_{2x} \, \sin u + 2 u_{3x} \, \cos u ] 
\nonumber \\
\!&\!&\! + [-2 u_{3x} \cos u - 6  u_x \, u_{2x} \sin u + 2  u_{3x} \cos u ] 
\nonumber \\
\!&\!\equiv \!&\! 0.
\end{eqnarray}
Similarly, for component $v$ (also of order $2)$ one readily verifies that
${\cal L}^{(0)}_{v(x)} (f) \equiv 0.$
%
\vskip 2pt
\noindent
{\bf Example}:
As an example in 2D, one can readily verify that 
$f =  u_x v_y - u_{2x} v_y - u_y v_x + u_{xy} v_x $ from 
(\ref{divergenceuv}) is a divergence.
Applying (\ref{zeroeulerscalaruxy}) to $f$ for each component of 
${\bf u} = (u,v)$ gives
${\cal L}^{(0,0)}_{u(x,y)} (f) \equiv 0$ and 
${\cal L}^{(0,0)}_{v(x,y)} (f) \equiv 0.$
\subsection{Continuous Higher Euler Operators}
\label{highereulercont}
%
%
To compute ${\bf F} = {\rm Div}^{-1}(f)$ or, in the 1D case
$F = {\rm D}_x^{-1}(f) = \int f \,dx,$ we need higher-order versions of the
variational derivative, called {\it higher Euler operators}.
The general formulas are given in \cite[p.\ 367]{PObook93}. 
With applications in mind, we restrict ourselves to the 1D, 2D, and 3D cases.
\vskip 5pt
\noindent
%
%
\noindent
{\bf Definition}:
The {\it higher Euler operators} in 1D (with variable $x$) are 
\begin{equation}
\label{highereulervectorux}
{\cal L}^{(i)}_{{\bf u}(x)} = 
\sum_{k=i}^{\infty} {k \choose i} (-{\rm D}_x)^{k-i} 
\frac{\partial }{\partial {\bf u}_{kx}}, 
\end{equation}
where ${k \choose i}$ is the binomial coefficient. 
Note that the higher Euler operator for $i=0$ matches the variational 
derivative in (\ref{zeroeulerscalarux}).
The explicit formulas for the first three higher Euler operators 
(for component $u$ and variable $x)$ are
\begin{eqnarray}
\label{highereulerscalarux}
{\cal L}^{(1)}_{u(x)} 
\!\!&\!=\!&\!\!
\frac{\partial }{\partial u_x} 
- 2{\rm D}_x     \frac{\partial }{\partial u_{2x}}
+ 3{\rm D}_{x}^2 \frac{\partial }{\partial u_{3x}} 
- 4{\rm D}_{x}^3 \frac{\partial }{\partial u_{4x}} + \cdots ,
\\
%
{\cal L}^{(2)}_{u(x)} 
\!\!&\!=\!&\!\!
\frac{\partial }{\partial u_{2x}} 
- 3{\rm D}_x \frac{\partial }{\partial u_{3x}}
+ 6{\rm D}_{x}^2 \frac{\partial }{\partial u_{4x}}
- 10{\rm D}_{x}^3 \frac{\partial }{\partial u_{5x}} + \cdots ,
\\
%
%
%
{\cal L}^{(3)}_{u(x)} 
\!&\!=\!&\!
\frac{\partial }{\partial u_{3x}} 
- 4{\rm D}_x \frac{\partial }{\partial u_{4x}}
+ 10{\rm D}_{x}^2 \frac{\partial }{\partial u_{5x}}
- 20{\rm D}_{x}^3 \frac{\partial }{\partial u_{6x}} 
+ \cdots .
\end{eqnarray}
%
%
%
%
\noindent
{\bf Definition}:
The {\it higher Euler operators} in 2D (with variables $x,y$) are given by
\begin{equation}
\label{highereulervectoruxy}
{\cal L}^{(i_{x},i_{y})}_{{\bf u}(x,y)} =
\sum_{k_{x}=i_{x}}^{\infty} \sum_{k_{y}=i_{y}}^{\infty}
{k_{x} \choose i_{x}} {k_{y} \choose i_{y}} 
(-{\rm D}_x)^{k_{x}-i_{x}}  (-{\rm D}_y)^{k_{y}-i_{y}} 
\frac{\partial }{\partial {\bf u}_{k_{x}x \, k_{y}y}} .
\end{equation}
%
%
Note that the higher Euler operator for $i_{x}\!=\!i_{y}\!=\!0$ matches 
the variational derivative in (\ref{zeroeulerscalaruxy}).
The first higher Euler operators (for component $u$ and variables $x$ 
and $y$) are
\begin{eqnarray}
\label{highereulerscalaruxy}
{\cal L}^{(1,0)}_{u(x,y)} 
\!\!&\!=\!&\!\!
\frac{\partial }{\partial u_x} 
\!-2 {\rm D}_x     \frac{\partial }{\partial u_{2x}}
\!-   {\rm D}_y     \frac{\partial }{\partial u_{xy}}
\!+3 {\rm D}_{x}^2 \frac{\partial }{\partial u_{3x}} 
\!+2 {\rm D}_{x} {\rm D}_{y} \frac{\partial }{\partial u_{2xy}} 
\!+   {\rm D}_{y}^2 \frac{\partial }{\partial u_{x2y}} 
\!- \cdots ,
\\
{\cal L}^{(0,1)}_{u(x,y)} 
\!\!&\!=\!&\!\!
\frac{\partial }{\partial u_y} 
\!-2 {\rm D}_y     \frac{\partial }{\partial u_{2y}}
\!-   {\rm D}_x    \frac{\partial }{\partial u_{yx}}
\!+3 {\rm D}_{y}^2 \frac{\partial }{\partial u_{3y}} 
\!+2 {\rm D}_{x} {\rm D}_{y} \frac{\partial }{\partial u_{x2y}} 
\!+   {\rm D}_{x}^2 \frac{\partial }{\partial u_{2xy}} 
\!- \cdots ,
\\
{\cal L}^{(1,1)}_{u(x,y)} 
\!\!&\!=\!&\!\!
\frac{\partial }{\partial u_{xy}} 
\!-2 {\rm D}_x \frac{\partial }{\partial u_{2xy}}
\!-2 {\rm D}_{y} \frac{\partial }{\partial u_{x2y}}
\!+3 {\rm D}_{x}^2 \frac{\partial }{\partial u_{3xy}} 
\!+4 {\rm D}_{x} {\rm D}_{y} \frac{\partial }{\partial u_{2x2y}} 
\!+ \cdots ,
\\
{\cal L}^{(2,1)}_{u(x,y)} 
\!&\!=\!&\!
\frac{\partial }{\partial u_{2xy}} 
\!-3 {\rm D}_x \frac{\partial }{\partial u_{3xy}}
\!-2 {\rm D}_{y} \frac{\partial }{\partial u_{2x2y}}
\!+6 {\rm D}_{x}^2 \frac{\partial }{\partial u_{4xy}} 
\!+3 {\rm D}_{y}^2 \frac{\partial }{\partial u_{2x3y}} 
-\! \cdots .
\end{eqnarray}
%
%
%
%
\vskip 5pt
\noindent
{\bf Definition}:
The {\it higher Euler operators} in 3D (with variables $x,y,z$) are 
\begin{equation}
\label{eulerhighervectoruxyz}
{\cal L}^{(i_{x},i_{y},i_{z})}_{{\bf u}(x,y,z)} \!\!=\!\!
\sum_{k_{x}=i_{x}}^{\infty}\!
\sum_{k_{y}=i_{y}}^{\infty} \!
\sum_{k_{z}=i_{z}}^{\infty} \!
{k_{x} \choose i_{x}} \!
{k_{y} \choose i_{y}}\!
{k_{z} \choose i_{z}} \!
(-{\rm D}_x)^{k_{x}-i_{x}} \!
(-{\rm D}_y)^{k_{y}-i_{y}} \!
(-{\rm D}_z)^{k_{z}-i_{z}} \!
\frac{\partial }{\partial {\bf u}_{k_{x}x \, k_{y}y \, k_{z}z}}.
\end{equation}
The higher Euler operator for $i_{x}\!=\!i_{y}\!=\!i_{z}\!=0\!$ matches 
the variational derivative given in (\ref{zeroeulerscalaruxyz}).
\subsection{Continuous Homotopy Operator}
\label{homotopyoperatorscont}
%
%
We now discuss the homotopy operator which will allow us to reduce the 
computation of ${\bf F} = {\rm Div}^{-1}(f)$ 
(or in the 1D case, $F = {\rm D}_x^{-1}(f) = \int f \, dx ) $
to a single integral with respect to an auxiliary variable 
denoted by $\lambda$ (not to be confused with $\lambda$ in 
Section~\ref{definitionsPDEs}).
Hence, the homotopy operator circumvents integration by parts and 
reduces the inversion of the total divergence operator, ${\rm Div},$ 
to a problem of single-variable calculus. 
The homotopy operator is given in explicit form, which makes it easier to 
implement in CAS.
To keep matters transparent, we present the formulas of the homotopy operator 
in 1D, 2D, and 3D.
%
%
\vskip 5pt
\noindent
{\bf Definition}:
The {\it homotopy operator} in 1D (with variable $x$) 
\cite[p.\ 372]{PObook93} is
\begin{equation}
\label{homotopyvectorux}
{\cal H}_{{\bf u}(x)}(f) = 
\int_{0}^{1} \sum_{j=1}^{N} I_{u_j}(f)
[\lambda {\bf u}] \, \frac{d \lambda}{\lambda},
\end{equation}
where the integrand $I_{u_j}(f)$ is given by 
\begin{equation}
\label{integrandhomotopyvectorux}
I_{u_j}(f) = \sum_{i=0}^{\infty} {\rm D}_{x}^i 
\left( u_j \, {\cal L}^{(i+1)}_{u_j(x)} (f) \right).
\end{equation}
The integrand involves the 1D higher Euler operators in 
(\ref{highereulervectorux}).
In (\ref{homotopyvectorux}), $N$ is the number of dependent variables and 
$I_{u_j}(f)[\lambda {\bf u}]$ means that in 
$I_{u_j}(f)$ one replaces ${\bf u}(x) \rightarrow \lambda {\bf u}(x), \, 
{\bf u}_x(x) \rightarrow \lambda {\bf u}_x(x),\, {\rm etc.}$
\vskip 5pt
\noindent
Given an exact function $f,$ the question how to compute 
$F = {\rm D}_x^{-1}(f) = \int f \, dx$ is then answered by the following 
theorem \cite[p.\ 372]{PObook93}.
\vskip 2pt
\noindent
{\bf Theorem}:
For an exact function $f,$ one has 
$F = {\cal H}_{{\bf u}(x)}(f).$
\vskip 3pt
\noindent
Thus, in the 1D case, applying the homotopy operator 
(\ref{homotopyvectorux}) allows one to bypass integration by parts.
A clever argument why the homotopy operator actually works is given in 
\cite[p.\ 582]{SAandGB02b}.
As an experiment, one can start from a function ${\tilde F},$ 
compute $f = {\rm D}_x {\tilde F},$ subsequently compute 
$F = {\cal H}_{{\bf u}(x)}(f),$ and finally verify that 
$F - {\tilde F}$ is a constant.
\vskip 4pt
\noindent
{\bf Example}:
Using (\ref{fcontinuous}), we show how the homotopy operator 
(\ref{homotopyvectorux}) is applied.
For a system with $N=2$ components, $(u_1,u_2)=(u,v),$ the homotopy operator 
formulas are
\begin{equation}
\label{homotopyuvx}
{\cal H}_{{\bf u}(x)} (f) = 
\int_{0}^{1} \left( 
I_{u}(f)[\lambda {\bf u}] + 
I_{v}(f)[\lambda {\bf u}] \right) 
\, \frac{d \lambda}{\lambda},
\end{equation}
with
\begin{equation}
\label{integrandhomotopyuvx}
I_{u}(f) = \sum_{i=0}^{\infty} {\rm D}_{x}^i 
\left( u \, {\cal L}^{(i+1)}_{u(x)} (f) \right)
\quad {\rm and} \quad
I_{v}(f) = \sum_{i=0}^{\infty} {\rm D}_{x}^i 
\left( v \, {\cal L}^{(i+1)}_{v(x)} (f) \right).
\end{equation}
These sums have only finitely many non-zero terms. 
For example, the sum in $I_{u}(f)$ terminates at $p-1$ where $p$ is the 
order of $u$. 
Take, for example, 
$f = 3 \, u_x \, v^2 \, \sin u - u_x^3 \, \sin u - 6 \, v \, v_x \, 
\cos u + 2 \, u_x \, u_{2x} \, \cos u + 8 \, v_x v_{2x}. $ 
%
%
First, we compute
\begin{eqnarray} 
\label{Iuforf}
I_{u}(f) &=& 
u {\cal L}^{(1)}_{u(x)} (f) + 
{\rm D}_x \left( u {\cal L}^{(2)}_{u(x)} (f) \right) 
\nonumber \\
&=& u \frac{\partial f}{\partial u_x} 
    - 2 u {\rm D}_x \left( \frac{\partial f}{\partial u_{2x}} \right)
    + {\rm D}_x \left( u \frac{\partial f}{\partial u_{2x}} \right) 
\nonumber \\
&=& 3 u v^2 \sin u - u u_x^2 \sin u + 2 u_x^2 \cos u.
\end{eqnarray} 
\noindent 
Next, 
\begin{eqnarray} 
\label{Ivforf}
I_{v}(f) &=& 
v {\cal L}^{(1)}_{v(x)} (f) + 
{\rm D}_x \left( v {\cal L}^{(2)}_{v(x)} (f) \right) 
\nonumber \\
&=& v \frac{\partial f}{\partial v_x} 
    - 2 v {\rm D}_x \left( \frac{\partial f}{\partial v_{2x}}  \right)
    + {\rm D}_x \left( v \frac{\partial f}{\partial v_{2x}} \right) 
\nonumber \\
&=& - 6 v^2 \cos u + 8 v_x^2.
\end{eqnarray} 
Formula (\ref{homotopyuvx}) reduces to an integral with respect to 
$\lambda:$
%
%
\begin{eqnarray}
\!\!\!\!\! F &\!\!=\!\!& {\cal H}_{{\bf u}(x)} (f) =
\int_0^1\!\left( I_{u} (f)[\lambda {\bf u}] 
+ I_{v} (f)[\lambda {\bf u}] \right) \frac{d\lambda}{\lambda} 
\nonumber \\ 
&\!\!=\!\!& \int_0^1\!\left( 3 \lambda^2 u v^2 \sin(\lambda u) 
  \!-\! \lambda^2 u u_x^2 \sin(\lambda u) 
  \!+\! 2 \lambda u_x^2 \cos(\lambda u) 
\!-\! 6 \lambda v^2 \cos(\lambda u) \!+\! 8 \lambda v_x^2 
\right) d\lambda  
\nonumber \\ 
& \!\!=\!\!& 4 v_x^2 + u_x^2  \cos u - 3 v^2  \cos u .
\end{eqnarray}
%
%
%
We now turn to the problem of inverting the ${\rm Div}$ operator using
the homotopy operator.
\vskip 4pt
\noindent
{\bf Definition}:
We define the {\it homotopy operator} in 2D (variables $x,y)$ 
through its two components 
$( {\cal H}^{(x)}_{{\bf u}(x,y)}(f), {\cal H}^{(y)}_{{\bf u}(x,y)}(f) ).$
The $x$-component of the operator is given by
\begin{equation}
\label{homotopyvectoruxycompx}
{\cal H}^{(x)}_{{\bf u}(x,y)} (f) = 
\int_{0}^{1} \sum_{j=1}^{N} I_{u_j}^{(x)} (f)
[\lambda {\bf u}] \, \frac{d \lambda}{\lambda},
\end{equation}
with $I_{u_j}^{(x)} (f)$ given by  
\begin{equation}
\label{integrandhomotopyvectoruxycompx}
I_{u_j}^{(x)}(f) = 
\sum_{i_x=0}^{\infty} \sum_{i_y=0}^{\infty} 
\left( \frac{1+i_x}{1+i_x+i_y} \right)
{\rm D}_{x}^{i_x} {\rm D}_{y}^{i_y} 
\left( u_j \, {\cal L}^{(1+i_x,i_y)}_{u_j(x,y)} (f) \right).
\end{equation}
Analogously, the $y$-component is given by
\begin{equation}
\label{homotopyvectoruxycompy}
{\cal H}^{(y)}_{{\bf u}(x,y)} (f) = 
\int_{0}^{1} \sum_{j=1}^{N} I_{u_j}^{(y)} (f) 
[\lambda {\bf u}] \, \frac{d \lambda}{\lambda},
\end{equation}
with 
\begin{equation}
\label{integrandhomotopyvectoruxycompy}
I_{u_j}^{(y)} (f) = 
\sum_{i_x=0}^{\infty} \sum_{i_y=0}^{\infty} 
\left( \frac{1+i_y}{1+i_x+i_y} \right)
{\rm D}_{x}^{i_x} {\rm D}_{y}^{i_y}
\left( u_j \, {\cal L}^{(i_x,1+i_y)}_{u_j(x,y)} (f) \right).
\end{equation}
These integrands involve the 2D higher Euler operators in 
(\ref{highereulervectoruxy}).
\vskip 5pt
\noindent
After verification that $f$ is a divergence, 
the question how to compute 
${\bf F} = (F_1,F_2) = {\rm Div}^{-1}(f)$ 
is then answered by the following theorem.
\vskip 2pt
\noindent
{\bf Theorem}:
If $F$ is a divergence, then
${\bf F} = (F_1, F_2) = {\rm Div}^{-1}(f) = 
({\cal H}^{(x)}_{{\bf u}(x,y)} (f), {\cal H}^{(y)}_{{\bf u}(x,y)} (f) ).$
\vskip 3pt
\noindent
The superscript $(x)$ in ${\cal H}^{(x)}(f)$ reminds us that we are 
computing the $x$-component of ${\bf F}.$
%
%
%
%
As a test, one can start from any vector ${\tilde{\bf F}}$ and 
compute $f = {\rm Div} \, {\tilde {\bf F}}.$ 
Next, compute 
${\bf F} = (F_1, F_2) = 
({\cal H}^{(x)}_{{\bf u}(x,y)} (f), {\cal H}^{(y)}_{{\bf u}(x,y)} (f) )$
and, finally, verify that ${\tilde{\bf F}} - {\bf F}$ is divergence free.
\vskip 4pt
\noindent
{\bf Example}:
Using (\ref{divergenceuv}), we show how the application of the 2D homotopy 
operator leads to (\ref{vectorfordivergenceuv}), up to a divergence 
free vector. 
Consider $f = u_x v_y - u_{2x} v_y - u_y v_x + u_{xy} v_x$, 
which is easily verified to be a divergence. 
In order to determine ${\rm Div}^{-1}(f)$, we calculate
\begin{eqnarray}
I_u^{(x)}(f) & = & u {\cal L}_{u(x,y)}^{(1,0)}(f)
+ {\rm D}_x \left( u{\cal L}_{u(x,y)}^{(2,0)}(f) \right)
+ \frac{1}{2}{\rm D}_y \left( u{\cal L}_{u(x,y)}^{(1,1)}(f) \right)
\nonumber \\
&=& u \left(\frac{\partial f}{\partial u_x}
- 2{\rm D}_x \frac{\partial f}{\partial u_{2x}}
- {\rm D}_y \frac{\partial f}{\partial u_{xy}}\right)
+ {\rm D}_x \left(u \frac{\partial f}{\partial u_{2x}} \right)
+ \frac{1}{2}{\rm D}_y \left( u \frac{\partial f}{\partial u_{xy}} \right)
\nonumber \\
&=& u v_y + \frac{1}{2} u_y v_x - u_x v_y + \frac{1}{2} u v_{xy}. 
\end{eqnarray}
Similarly, 
\begin{equation}
I_v^{(x)}(f)= v {\cal L}_{v(x,y)}^{(1,0)}(f)
= v \frac{\partial f}{\partial v_x}=-u_y v+u_{xy}v . 
\end{equation}
Hence, 
\begin{eqnarray}
\label{applhomotop2Dxpart}
\!\!\!\!\! F_1 &\!\!=\!\!& {\cal H}^{(x)}_{{\bf u}(x,y)} (f) 
= \int_{0}^{1} \left( I_{u}^{(x)} (f) [\lambda {\bf u}] + 
I_{v}^{(x)} (f) [\lambda {\bf u}] \right) \, \frac{d \lambda}{\lambda}
\nonumber \\
&\!\!=\!\!& \int_0^1\!\lambda \left( 
u v_y + \frac{1}{2} u_y v_x - u_x v_y + \frac{1}{2} u v_{xy} 
- u_y v + u_{xy} v \right) \, d\lambda  
\nonumber \\ 
&\!\!=\!\!& 
\frac{1}{2} u v_y + \frac{1}{4} u_y v_x - \frac{1}{2} u_x v_y 
+ \frac{1}{4} u v_{xy}  - \frac{1}{2} u_y v + \frac{1}{2} u_{xy} v.
\end{eqnarray}
Without showing the details, one computes in an analogous fashion
\begin{eqnarray}
\label{applhomotop2Dypart}
\!\!\!\!\! F_2 &\!\!=\!\!& {\cal H}^{(y)}_{{\bf u}(x,y)} (f) 
= \int_{0}^{1} \left( 
I_{u}^{(y)} (f) [\lambda {\bf u}] + 
I_{v}^{(y)} (f) [\lambda {\bf u}] \right) \, \frac{d \lambda}{\lambda}
\nonumber \\
&\!\!=\!\!& \int_0^1\!\lambda \left( 
- u v_x - \frac{1}{2} u v_{2x} + \frac{1}{2} u_x v_x \right)
+ \lambda \left( u_x v - u_{2x} v \right) \, d\lambda  
\nonumber \\ 
&\!\!=\!\!& 
- \frac{1}{2} u v_x - \frac{1}{4} u v_{2x} + \frac{1}{4} u_x v_x 
+ \frac{1}{2} u_x v- \frac{1}{2} u_{2x} v.
\end{eqnarray}
One can readily verify that the resulting vector 
\begin{equation}
\label{vectorf2d}
{\bf F} = \left(
\begin{array}{c}
 F_1 \\
 F_2 \end{array} \right) = 
\left(\!
\begin{array}{c}
\frac{1}{2} u v_y + \frac{1}{4} u_y v_x - \frac{1}{2} u_x v_y 
+ \frac{1}{4} u v_{xy}  - \frac{1}{2} u_y v + \frac{1}{2} u_{xy} v \\
- \frac{1}{2} u v_x - \frac{1}{4} u v_{2x} + \frac{1}{4} u_x v_x 
+ \frac{1}{2} u_x v- \frac{1}{2} u_{2x} v \end{array}
\! \right)
\end{equation}
differs from ${\tilde{\bf F}} = (u v_y - u_x v_y , -u v_x + u_x v_x)$ by a 
divergence free vector.
\newline
\noindent
%
The generalization of the homotopy operator to 3D is straightforward.
\vskip 2pt
\noindent
{\bf Definition}:
The {\it homotopy operator} in 3D (with variables $x,y,z)$ is 
\newline
$( {\cal H}^{(x)}_{{\bf u}(x,y,z)} (f), {\cal H}^{(y)}_{{\bf u}(x,y,z)} (f), 
{\cal H}^{(z)}_{{\bf u}(x,y,z)} (f)).$
By analogy with (\ref{homotopyvectoruxycompx}), the $x$-component is
\begin{equation}
\label{homotopyvectoruxyzcompx}
{\cal H}^{(x)}_{{\bf u}(x,y,z)} (f) = 
\int_{0}^{1} \sum_{j=1}^{N} I_{u_j}^{(x)} (f)
[\lambda {\bf u}] \, \frac{d \lambda}{\lambda},
\end{equation}
with,
\begin{equation}
\label{integrandhomotopyvectoruxyzcompx}
I_{u_j}^{(x)} (f) = 
\sum_{i_x=0}^{\infty} \sum_{i_y=0}^{\infty} \sum_{i_z=0}^{\infty} 
\left( \frac{1+i_x}{1+i_x+i_y+i_z} \right)
{\rm D}_{x}^{i_x} {\rm D}_{y}^{i_y} {\rm D}_{z}^{i_z} 
\left( u_j \, {\cal L}^{(1+i_x,i_y,i_x)}_{u_j(x,y,z)} (f) \right).
\end{equation}
The $y$ and $z$-operators are defined analogously.
The integrands involve the 3D higher Euler operators in 
(\ref{eulerhighervectoruxyz}).
By analogy with the 2D case the following theorem holds.
\vskip 2pt
\noindent
{\bf Theorem}:
Given a divergence $f$ one has 
${\bf F} \!=\! {\rm Div}^{-1}(f) \!=\!
( {\cal H}^{(x)}_{{\bf u}(x,y,z)} (f),
{\cal H}^{(y)}_{{\bf u}(x,y,z)} (f),
{\cal H}^{(z)}_{{\bf u}(x,y,z)} (f) ).$
\section{Removing Divergences and Equivalent Terms}
\label{algorithm}
We present an algorithm to remove divergences and equivalent terms in 
order to make our computation of the densities simpler.
\vskip 3pt
\noindent
{\bf Definition}: 
Two scalar differential functions, $f^{(1)}$ and $f^{(2)},$ are 
{\it equivalent} if and only if they differ by the divergence of some vector 
${\bf V},$ i.e.\ $f^{(1)} \sim f^{(2)}$ if and only if 
$f^{(1)} - f^{(2)} = {\rm Div} \, {\bf V}.$
\vskip 1pt
\noindent
Obviously, if a scalar expression is equivalent to zero, then it is a 
divergence.
\vskip 2pt
\noindent
{\bf Example}: 
Functions $f^{(1)} = u u_{2x}$ and $f^{(2)} = - u_x^2$ are equivalent 
because $f^{(1)} - f^{(2)} = u u_{2x} + u_x^2 = {\rm D}_x (u u_x).$
Using (\ref{zeroeulerscalarux}), note that
$v_1 = {\cal L}_{u(x)}^{(0)} (u u_{2x}) = 2 u_{2x}$ and 
$v_2 = {\cal L}_{u(x)}^{(0)} (-u_x^2) = 2 u_{2x}$ are equal.
Also, $f = u_{4x} = {\rm D}_x (u_{3x})$ is a divergence and, 
as expected, $v_3 = {\cal L}_{u(x)}^{(0)} (u_{4x}) = 0.$
\vskip 3pt
\noindent
{\bf Example}: 
In the 2D case, $f^{(1)} = (u_x - u_{2x}) v_y$ and 
$f^{(2)} = (u_y - u{xy} ) v_x $ are equivalent since 
$f^{(1)} - f^{(2)} = u_x v_y - u_{2x} v_y - u_y v_x + u_{xy} v_x  
= {\rm Div} \, ( u v_y - u_x v_y , - u v_x + u_x v_x).$
Using (\ref{zeroeulerscalaruxy}), note that
${\bf v}_1 = {\cal L}_{{\bf u}(x,y)}^{(0)} (f^{(1)}) =
{\bf v}_2 = {\cal L}_{{\bf u}(x,y)}^{(0)} (f^{(2)}) = 
(-v_{xy} - v_{xxy}, - u_{xy} + u_{xxy}).$
%
%
\vskip 2pt
\noindent
To remove divergences and equivalent terms we use the following algorithm.
\vskip 3pt
\noindent
%
%
%
\vfill
\newpage
%
\noindent
{\bf Algorithm}: {\sc Remove-Divergences-And-Equivalent-Terms}$({\cal R})$
\begin{verse}
 {\sc /* Given is list} ${\cal R}$ of monomial 
 differential functions {\sc */} \\
 {\sc /* Initialize two new lists} ${\cal S, B}$ {\sc */} \\
 ${\cal S} \leftarrow \emptyset$ \\
 ${\cal B} \leftarrow \emptyset$ \\
 {\sc /* Find first member of} ${\cal S}$ {\sc */} \\
 {\bf for} each term $t_i \in {\cal R}$ 
 \vspace{-2.75mm}
  \begin{verse}
     {\bf do} ${\bf v}_i \leftarrow {\cal L}_{{\bf u}({\bf x})}^{(0)} (t_i)$ 
    \begin{verse}
       {\bf if} ${\bf v}_i \ne {\bf 0}$ 
            \begin{verse}
               {\bf then} ${\cal S} \leftarrow \{ t_i \}$  \\
               $\quad\quad\;\; {\cal B} \leftarrow \{ {\bf v}_i \}$ \\
               $\quad\quad\;\; ${\bf break} \\
               {\bf else} discard $t_i$ and ${\bf v}_i$ 
            \end{verse} 
    \end{verse}
  \end{verse}
 \vspace{-2.75mm}
 {\sc $\!\!\!\!\!\!\!\!$/* Find remaining members of} ${\cal S}$ {\sc */} \\
 {\bf for} each term $t_j \in {\cal R} \setminus \{t_1, t_2, \cdots ,t_i\}$
 \vspace{-2.75mm}
   \begin{verse}
    {\bf do} ${\bf v}_j \leftarrow {\cal L}_{{\bf u}({\bf x})}^{(0)} (t_j)$ 
      \begin{verse}
       {\bf if} ${\bf v}_j \ne {\bf 0}$ 
         \begin{verse}
          {\bf then if} ${\bf v}_j \not\in {\rm Span}({\cal B})$
            \begin{verse}
             $\quad${\bf then} ${\cal S}\leftarrow {\cal S}\cup\{t_j\}$\\
             $\quad\quad\quad\;\; 
                           {\cal B} \leftarrow {\cal B} \cup \{{\bf v}_j\}$\\
             $\quad\,${\bf else} discard $t_j$ and ${\bf v}_j$ 
            \end{verse}         
         \end{verse}
      \end{verse}  
   \end{verse}
$\!\!\!\!\!\!\!\!\!${\bf return} ${\cal S}$ \\
{\sc /* List} ${\cal S}$ {\sc is free of divergences and equivalent terms */} 
\end{verse}
\vskip 1pt
\noindent
{\bf Example}:
Using the above algorithm, we remove divergences and equivalent terms in 
${\cal R} = \{u^3, u^2 v, u v^2, v^3, u_x^2, u_x v_x, v_x^2, u u_{2x}, 
u_{2x} v, u v_{2x}, v v_{2x}, u_{4x}, v_{4x} \}. $
Since 
${\bf v}_1 = {\cal L}_{{\bf u}({\bf x})}^{(0)} (u^3) = (3 u^2, 0) \ne (0,0)$
we have ${\cal S} = \{ t_1 \} = \{ u^3 \}$ and 
${\cal B} = \{ {\bf v}_1 \} = \{ (3 u^2, 0) \}.$ 
The first for loop is halted and the second for loop starts.
Next, ${\bf v}_2 = {\cal L}_{{\bf u}({\bf x})}^{(0)} (u^2 v) = 
(2 u v, u^2) \ne (0,0).$
We verify that ${\bf v}_1$ and ${\bf v}_2$ are independent and
update the sets resulting in 
${\cal S} = \{ t_1, t_2 \} = \{u^3, u^2 v \}$ 
and 
${\cal B} = \{ {\bf v}_1, {\bf v}_2 \} = \{ (3 u^2, 0), (2 u v, u^2) \}.$

Proceeding in a similar fashion, since the first seven terms
are indeed independent, we have  ${\cal S} = \{ t_1, t_2, \cdots, t_7 \}$
and ${\cal B} = \{ {\bf v}_1, {\bf v}_2, \cdots, {\bf v}_7 \} = 
\{ (3 u^2, 0), (2 u v, u^2), \cdots, (0, -2 v_{2x}) \}.$
\newline
For $t_8 = u u_{2x}$ we compute 
${\bf v}_8 = {\cal L}_{{\bf u}({\bf x})}^{(0)} (u u_{2x}) = (2 u_{2x}, 0)$ 
and verify that ${\bf v}_8 = - {\bf v}_5.$ 
So, ${\bf v}_8 \in {\rm Span}({\cal B})$ and $t_8$ and ${\bf v}_8$ are
discarded (i.e.\ {\it not} added to the respective sets).
For similar reasons, $t_{9}, t_{10},$ and $t_{11}$ 
as well as ${\bf v}_9, {\bf v}_{10},$ and ${\bf v}_{11}$ are discarded. 
The terms $t_{12} = u_{4x}$ and $t_{13} = v_{4x}$ are discarded because
${\bf v}_{12} = {\bf v}_{13} = (0,0).$
So, ${\cal R}$ is replaced by 
${\cal S} = \{ u^3, u^2 v, u v^2, v^3, u_x^2, u_x v_x, v_x^2 \}$ 
which is free of divergences and equivalent terms. 

\section{Application: Conservation Laws of Nonlinear PDEs}
\label{applicationsPDEs}
As an application of the Euler and homotopy operators we show how 
to compute conserved densities and fluxes for the three PDEs in 
Section~\ref{continuousexamples}.
The first PDE illustrates the 1D case (one independent variable),
but it involves two dependent variables $u(x)$ and $v(x).$
The second PDE (again in 1D) has a transcendental nonlinearity which 
complicates the computation of conserved densities and fluxes 
\cite{RDandRBprsa77}. 
A third example illustrates the algorithm for a 2D case.

To compute conservation laws 
${\rm D}_{t} \, \rho + {\rm Div} \, {\bf J} = 0$ of systems of 
nonlinear PDEs, we use a direct approach.
First, we build the candidate density $\rho$ as a linear combination 
(with constant coefficients $c_i)$ of terms which are 
uniform in rank (with respect to the scaling symmetry of the PDE). 
It is of paramount importance that the candidate density is free of 
divergences and equivalent terms. 
If such terms were present, their coefficients could not be determined 
because such terms can be moved into the flux, ${\bf J}.$
To construct the shortest density, 
we will use the algorithm of Section~\ref{algorithm}.

Second, we evaluate ${\rm D}_t \rho$ on solutions of the PDE, thus removing 
all time derivatives from the problem.
The resulting expression (called $E)$ must be a divergence 
(of the as yet unknown flux). 
Thus, we set ${\cal L}_{{\bf u}({\bf x})}^{(0)} (E) \equiv 0.$ 
Setting the coefficients of like terms to zero leads to a linear system 
for the undetermined coefficients $c_i.$
In the most difficult case, such systems are parameterized by the constant 
parameters appearing in the given PDE. 
If so, a careful analysis of the eliminant (and solution branching) 
must be carried out. 
For each branch, the solution of the linear system is substituted into
$\rho$ and $E.$ 

Third, since $E = {\rm Div} \, {\bf J}$ we use the homotopy operator 
${\cal H}_{{\bf u}({\bf x})}$ to compute ${\bf J} = {\rm Div}^{-1}(E).$
The computations are carried out with our {\it Mathematica} packages 
\cite{WHwebsite04}.
\subsection{Conservation Laws for the Coupled KdV Equations}
\label{applckdv}
%
%
In (\ref{ckdvconslaw1}) through (\ref{ckdvconslaw4}) we gave the first 
four density-flux pairs.
As an example, we will compute density ${\rho}^{(4)}$ and associated flux 
$J^{(4)}.$ 

Recall that the weights for the cKdV equations are 
$W({\partial}/{\partial x}) = 1$ and $W(u) = W(v) = 2.$ 
The parameter $\beta$ has no weight. 
Hence, ${\rho}^{(4)}$ has rank 6. 
The algorithm has three steps:
\vskip 4pt
\noindent
{\bf Step 1}: {\bf Construct the form of the density}
\vskip 2pt
\noindent
Start from ${\cal V} = \{ u, v \},$ i.e.\ the list of dependent variables 
with weight.
Construct 
${\cal M} \!=\! \{ u^3, v^3, u^2 v, u v^2, u^2, v^2, u v, u, v, 1 \},$
which contains all monomials of selected rank $6$ or less 
(without derivatives).
Next, for each monomial in ${\cal M},$ introduce the correct number of 
$x$-derivatives so that each term has rank 6.
For example, 
\begin{eqnarray}
\label{buildingblocksckdv}
{{\partial}^2 u^2\over \partial{x^2}} 
\!&\!=\!&\! 2 u_x^2 + 2 u u_{2x}, \quad
{{\partial}^2 v^2\over \partial{x^2}} = 2 v_x^2 + 2 v v_{2x}, \quad
{{\partial}^2 ( u v )\over \partial{x^2}}  = u_{2x} v + 2 u_x v_x + u v_{2x}, 
\nonumber \\
{{\partial}^4 u\over \partial{x^4}}  \!&\!=\!&\! u_{4x}, \quad
{{\partial}^4 v\over \partial{x^4}}  = v_{4x}, \quad
{{\partial}^6 1\over \partial{x^6}}  = 0.
\end{eqnarray}
Ignore the highest-order terms (typically the last terms) in each 
of the right hand sides of (\ref{buildingblocksckdv}). 
Augment ${\cal M}$ with the remaining terms, after stripping off 
numerical factors, to get
${\cal R} = \{ u^3, u^2 v, u v^2, v^3, u_x^2, u_x v_x, v_x^2, u_{2x} v \},$
where the 8 terms are listed by increasing order.

Note that keeping all terms in (\ref{buildingblocksckdv}) would have 
resulted in the list ${\cal R}$ 
(with 13 terms) given in the example at the end of Section~\ref{algorithm}. 
As shown, the algorithm would reduce ${\cal R}$ to 7 terms.

%
%
%

Use the algorithm of Section~\ref{algorithm}, to replace ${\cal R}$ 
by ${\cal S} = \{ u^3, u^2 v, u v^2, v^3, u_x^2, u_x v_x, v_x^2 \}.$ 
Linearly combine the terms in ${\cal S}$ with constant coefficients 
to get the shortest candidate density:
\begin{equation}
\label{candidaterhockdv}
\rho = c_1 u^3 + c_2 u^2 v + c_3 u v^2 + c_4 v^3 + c_5 u_x^2 
+ c_6 u_x v_x + c_7 v_x^2. 
\end{equation}
\vskip 4pt
\noindent
{\bf Step 2}: {\bf Determine the constants $c_i$}
\vskip 2pt
\noindent
Compute 
\begin{eqnarray}
\label{lastEckdv}
E \!=\! {\rm D}_t \rho 
&\!\!=\!\!&\! 
\frac{\partial \rho}{\partial t} + \rho^{\prime}({\bf u}) [{\bf F}] 
\!=\! \frac{\partial \rho}{\partial u} u_t 
+ \frac{\partial \rho}{\partial u_{x}} u_{tx}
+ \frac{\partial \rho}{\partial v} v_t 
+ \frac{\partial \rho}{\partial v_{x}} v_{tx} 
\nonumber \\
&\!\!=\!\!&\! 
(3 c_1 u^2 + 2 c_2 u v + c_3 v^2) u_t
+ (2 c_5 u_x + c_6 v_x) u_{tx}
+ (c_2 u^2 + 2 c_3 u v + 3 c_4 v^2) v_t
\nonumber \\
&&\!+ (c_6 u_x + 2 c_7 v_x) v_{tx}.
\end{eqnarray}
Replace $u_t, v_t, u_{tx}$ and $v_{tx}$ from (\ref{ckdv}) to obtain
\begin{eqnarray}
\label{expressionE}
E &\!\!=\!\!&\! 
(3 c_1 u^2 + 2 c_2 u v + c_3 v^2) (6 \beta u u_x - 6 v v_x + \beta u_{3x}) 
+ (2 c_5 u_x + c_6 v_x) (6 \beta u u_x - 6 v v_x + \beta u_{3x})_x 
\nonumber \\
&& - (c_2 u^2 + 2 c_3 u v + 3 c_4 v^2) (3 u v_x + v_{3x})
- (c_6 u_x + 2 c_7 v_x) (3 u v_x + v_{3x})_x.
\end{eqnarray}
Since $E \!=\! {\rm D}_t \, \rho \!=\! -{\rm D}_x J,$ the expression $E$ 
must be exact.
Therefore, apply the variational derivative (\ref{zeroeulerscalarux}) 
and require that 
${\cal L}^{(0)}_{u(x)} (E) \equiv 0$ and 
${\cal L}^{(0)}_{v(x)} (E) \equiv 0.$ 
Group like terms and set their coefficients equal to zero to obtain the 
following (parameterized) linear system for the unknown coefficients 
$c_1$ through $c_7:$ 
\begin{eqnarray}
&&\!\!\!\!\!\!\!\!\!
(3 + 4 \beta) c_2 = 0, \;\; 3 c_1 + (1 + \beta) c_3 = 0, \;\; 
4 c_2 + 3 c_4 = 0,\;\; (1+\beta) c_3 - 6 c_5 = 0, 
\nonumber \\
&&\!\!\!\!\!\!\!\!\!
\beta (c_1 + 2 c_5) = 0, \;\;
\beta c_2 - c_6 = 0, \;\; (1 + \beta) c_6 = 0,\;\; c_4 + c_6 = 0, 
\\
&&\!\!\!\!\!\!\!\!\!
2 (1 + \beta) c_2 -  3(1 + 2 \beta) c_6 = 0, \;\;
2 c_2 - (1 + 6 \beta) c_6 = 0, \;\; 
\beta c_3 - 6 c_5 - c_7 = 0, \;\; c_3 + c_{7} = 0.
\nonumber
\end{eqnarray}
Investigate the eliminant of the system. 
In this example, there exists a solution for any $\beta \ne -1.$
Set $c_1 = 1$ and obtain 
\begin{equation}
\label{solcickdv}
c_1 = 1, \; c_2 = c_4 = c_6 = 0,\; c_3 = -\textstyle{\frac{3}{1+\beta}}, \; 
c_5 = -\frac{1}{2}, \; c_7 = \textstyle{\frac{3}{1+\beta}}.
\end{equation}
Substitute the solution into (\ref{candidaterhockdv}) and multiply by 
$1+\beta$ to get
\begin{equation}
\label{rhockdv}
\rho = (1 + \beta ) u^3 - 3 u v^2 
- \textstyle{\frac{1}{2}}(1+\beta) {u_x}^2 + 3 {v_x}^2,
\end{equation}
which is $\rho^{(4)}$ in (\ref{ckdvconslaw4}).
\vskip 4pt
\noindent
{\bf Step 3}: {\bf Compute the flux $J$}
\vskip 2pt
\noindent
Compute the flux corresponding to $\rho$ in (\ref{rhockdv}).
Substitute (\ref{solcickdv}) into (\ref{expressionE}), 
reverse the sign and multiply by $1+\beta,$ to get 
\begin{eqnarray}
\label{expressionEsimp}
E &\!\!=\!\!&\! 
18 \beta (1 + \beta) u^3 u_x
- 18 \beta u^2 v v_x
- 18 \beta u u_x v^2
+ 18 v^3 v_x 
- 6 \beta (1 + \beta) u_x^3
- 6 \beta (1 + \beta) u u_x u_{2x}
\nonumber \\
&& + 3 \beta (1 + \beta) u^2 u_{3x}
- 3 \beta v^2 u_{3x}
- 6 v_x v_{4x}
- \beta (1 + \beta) u_x u_{4x}
+ 6 u v u_{3x}
+ 6 (\beta - 2 ) u_x v_x^2
\nonumber \\
&& + 6 (1 + \beta) u_x v v_{2x}
- 18 u v_x v_{2x}
\end{eqnarray}
Apply (\ref{homotopyvectorux}) and (\ref{integrandhomotopyvectorux}) 
to (\ref{expressionEsimp}) to obtain
\begin{eqnarray}
\label{jckdv}
J \!&\!=\!&\! 
-\textstyle{\frac{9}{2}} \beta (1 + \beta ) u^4 
+ 9 \beta u^2 v^2 
- \textstyle{\frac{9}{2}} v^4
+ 6 \beta (1 + \beta ) u u_x^2 
- 3 \beta (1 + \beta ) u^2 u_{2x}
\nonumber \\
&& + 3 \beta v^2 u_{2x} 
- \textstyle{\frac{1}{2}} \beta (1 + \beta ) u_{2x}^2 
+ \beta (1 + \beta ) u_x u_{3x} 
- 6 \beta v u_x v_x 
\nonumber \\
&& + 12 u v_x^2 
- 6 u v v_{2x} 
- 3 v_{2x}^2 
+ 6 v_x v_{3x},
\end{eqnarray}
which is $J^{(4)}$ in (\ref{ckdvconslaw4j}).

The cKdV equations (\ref{ckdv}) are completely integrable if 
$\beta = \frac{1}{2}$ and admit conserved densities at every even rank.
\subsection{Conservation Laws for the sine-Gordon Equation}
\label{applsg}
Recall that the weights for the sine-Gordon equation (\ref{sinegordon0}) are 
$W(\frac{\partial}{\partial x}) = 1, W(u) = 0, W(v) = 1,$ and $W(\alpha) = 2.$
The first few (of infinitely many) densities and fluxes were given in 
(\ref{sinegordonconslaw1}) through (\ref{sinegordonconslaw4}). 
We show how to compute densities ${\rho}^{(1)}$ and ${\rho}^{(2)},$
both of rank 2, and 
their associated fluxes $J^{(1)}$ and $J^{(2)}.$ 

In contrast to the previous example, the candidate density will no longer
have {\it constant} undetermined coefficients $c_i$ but {\it functional} 
coefficients $h_i(u)$ which depend on the transcendental variable $u$ 
with weight zero \cite{PAthesis03}. 
To avoid having to solve PDEs, we tacitly assume that there is only 
{\it one} dependent variable with weight zero.
\vskip 4pt
\noindent
{\bf Step 1}: {\bf Construct the form of the density}
\vskip 2pt
\noindent
Augment the list of dependent variables with $\alpha$ 
(with non-zero weight) and replace $u$ by $u_x$ (since $W(u)=0).$
Hence, ${\cal V} = \{\alpha , u_x, v \}.$
Compute ${\cal R} = \{ \alpha, v^2, v^2, u_{2x}, u_x v, u_x^2 \}$ and
remove divergences and equivalent terms to get
${\cal S} = \{\alpha, v^2, u_x^2, u_x v \}.$ 
The candidate density
\begin{equation}
\label{candidaterhosg}
\rho = \alpha h_1(u) + h_2(u) v^2 + h_3(u) {u_x}^2 + h_4(u) u_x v, 
\end{equation}
with undetermined functional coefficients $h_i(u).$ 
\vskip 4pt
\noindent
{\bf Step 2}: {\bf Determine the functions $h_i(u)$}
\vskip 2pt
\noindent
Compute 
\begin{eqnarray}
\label{lastEsinegordon}
\!\!\!\!\!\!\!\!\!E \!=\! {\rm D}_t \rho &\!\!=\!\!&\! 
\frac{\partial \rho}{\partial t} + \rho^{\prime}({\bf u}) [{\bf F}] 
\!=\! \frac{\partial \rho}{\partial u} u_t 
+ \frac{\partial \rho}{\partial u_{x}} u_{tx} + 
\frac{\partial \rho}{\partial v} v_t 
\nonumber \\
%
%
%
&\!\!=\!\!&\! (\alpha h_1^{\prime} \!+\! v^2 h_2^{\prime} 
\!+\! u_x^2 h_3^{\prime} \!+\! u_x v h_4^{\prime}) v
\!+\! (2 u_x h_3 \!+\! v h_4 ) v_x \!+\!\!
(2 v h_2 \!+\! u_x h_4 ) (\alpha \sin(u) \!+\! u_{2x}).
\end{eqnarray}
where $h_i^{\prime}$ means $\textstyle{\frac{dh_i}{du}}.$
Since $E \!=\! {\rm D}_t \, \rho \!=\! -{\rm D}_x J,$ 
the expression $E$ must be exact.
Therefore, 
require that 
${\cal L}^{(0)}_{u(x)} (E) \equiv 0$ and 
${\cal L}^{(0)}_{v(x)} (E) \equiv 0.$ 
Set the coefficients of like terms equal to zero to get a mixed linear 
system of algebraic and ODEs: 
\begin{eqnarray}
&& h_2(u) - h_3(u) \!=\! 0, \quad h_2^{\prime}(u) \!=\! 0, \quad 
h_3^{\prime}(u) \!=\! 0, \quad h_4^{\prime}(u) \!=\! 0,  \quad
h_2^{\prime\prime}(u) \!=\! 0, \\ 
&& h_4^{\prime\prime}(u) = 0, \quad
2 h_2^{\prime}(u) - h_3^{\prime}(u) = 0, \quad 
2 h_2^{\prime\prime}(u) - h_3^{\prime\prime}(u) = 0, \\
&& h_1^{\prime}(u)  + 2 h_2(u) \sin u = 0, \quad 
h_1^{\prime\prime}(u) + 2 h_2^{\prime}(u) \sin u + 2  h_2(u) \cos u = 0. 
\end{eqnarray}
Solve the system \cite{PAthesis03} and substitute the solution
\begin{equation}
\label{solhisinegordon}
{h_1}(u) = 2 c_1  \cos u + c_3, \quad {h_2}(u) = {h_3}(u) = c_1, 
\quad {h_4}(u) = c_2,
\end{equation}
(with arbitrary constants $c_i)$ into (\ref{candidaterhosg}) to obtain
\begin{equation}
\label{rhosinegordon}
\rho = c_1 (2 \alpha  \cos u + v^2 + {u_x}^2) + c_2 u_x v + c_3 \alpha.
\end{equation}
\vskip 4pt
\noindent
{\bf Step 3}: {\bf Compute the flux $J$}
\vskip 2pt
\noindent
Compute the flux corresponding to $\rho$ in (\ref{rhosinegordon}).
Substitute (\ref{solhisinegordon}) into (\ref{lastEsinegordon}), to get
%
%
\begin{equation}
\label{Eevalsinegordon}
E = c_1 ( 2 u_{2x} v + 2 u_x v_x ) 
    + c_2 ( v v_x + u_x u_{2x} + \alpha u_x \sin u ) .
\end{equation}
%
Since $E \!=\! {\rm D}_t \, \rho \!=\! -{\rm D}_x J,\;$ one must integrate 
$f\!=\!-E.$
Applying (\ref{integrandhomotopyuvx}) yields
$I_{u} (f) \!=\! - 2 c_1 u_x v - c_2 (u_{x}^2 + \alpha u \sin u )$
and 
$I_{v} (f) \!=\! - 2 c_1 u_x v - c_2 v^2. $
Use formula (\ref{homotopyuvx}) to obtain
\begin{eqnarray}
J &=& {\cal H}_{{\bf u}(x)} (f) = 
\int_0^1 \left( I_{u} (f)[\lambda {\bf u}] 
+ I_{v} (f)[\lambda {\bf u}] \right) \; \frac{d\lambda}{\lambda} 
\nonumber \\ 
&=& - \int_0^1 
\left( 
4 c_1 \lambda u_x v + c_2 (\lambda u_x^2 + \alpha u \sin(\lambda u) 
  + \lambda v^2) \right) \; d\lambda  
\nonumber \\
&=& 
- c_1 (2 u_x v) - c_2 \left(
\frac{1}{2} v^2 + \frac{1}{2} u_x^2 - \alpha  \cos u \right) .
\label{fluxforEevalsinegordon}
\end{eqnarray}
Finally, split density (\ref{rhosinegordon}) and 
flux (\ref{fluxforEevalsinegordon}) into independent pieces 
(for $c_1$ and $c_2)$:
\begin{eqnarray}
\rho^{(1)} \!&\!=\!&\! 2 \alpha  \cos u + v^2 + {u_x}^2  \quad {\rm and} \quad
J^{(1)} = - 2 u_x v, \\
\rho^{(2)} \!&\!=\!&\! u_x v \quad {\rm and} \quad 
J^{(2)} = - \frac{1}{2} v^2 - \frac{1}{2} u_x^2 + \alpha  \cos u.
\end{eqnarray}
For $E$ in (\ref{Eevalsinegordon}), $J$ in 
(\ref{fluxforEevalsinegordon}) can easily be computed by hand 
\cite{PAthesis03}.
However, the computation of fluxes corresponding to densities of ranks 
$\ge 2$ is cumbersome and requires integration with the homotopy operator.
%
\subsection{Conservation Laws for the Shallow Water Wave Equations}
\label{applsww}
In contrast to the previous two examples, as far as we know, 
(\ref{sww}) is not completely integrable.
One cannot expect infinitely many conserved densities and fluxes 
(of different ranks).

The first few densities and fluxes were given in (\ref{swwconslaw}). 
We show how to compute densities 
${\rho}^{(1)}, {\rho}^{(3)}, {\rho}^{(4)},$ and ${\rho}^{(5)},$
which are of rank 3 under the following (choice for the) weights
\begin{equation}
\label{swwscale1}
W({\partial/\partial x}) \!=\! W({\partial/\partial y}) \!=\! 1,  
W(u) \!=\! W(v) \!=\! 1, W(\theta) \!=\! 1, W(h) \!=\! 1, W(\Omega) \!=\! 2.
\end{equation}
We will also compute the associated fluxes 
$J^{(1)}, J^{(3)}, J^{(4)},$ and $J^{(5)}.$ 

The fact that (\ref{sww}) is multi-uniform is advantageous. 
Indeed, one can use the invariance of (\ref{sww}) under one scale to 
construct the terms of $\rho,$ and, 
subsequently, use additional scale(s) to split $\rho$ into smaller densities.
This ``divide and conquer" strategy drastically reduces the complexity of 
the computations.
\vskip 4pt
\noindent
{\bf Step 1}: {\bf Construct the form of the density}
\vskip 2pt
\noindent
Start from ${\cal V} = \{ u, v, \theta, h, \Omega \},$ 
i.e.\ the list of variables {\em and} parameters with weights.
Use (\ref{swwscale1}) to construct
${\cal M} = 
\{ \Omega u, \Omega v, \cdots, u^3, v^3, \cdots, u^2 v, u v^2, \cdots,
u^2, v^2, uv, \cdots, u, v, \theta, h \},$
which has 38 monomials of rank 3 or less (without derivatives). 

The terms of rank 3 in ${\cal M}$ are left alone. 
To adjust the rank, differentiate each monomial of rank 2 in ${\cal M}$
with respect to $x$ ignoring the highest-order term. 
For example, in $\frac{du^2}{dx} = 2 u u_x,$ the term can be ignored
since it is a total derivative.
The terms $u_x v$ and $-u v_x$ are equivalent since 
$\frac{d (uv)}{dx} = u_x v + u v_x.$ 
Keep $u_x v.$
Likewise, differentiate each monomial of rank 2 in ${\cal M}$ with 
respect to $y$ and ignore the highest-order term. 

Produce the remaining terms for rank 3 by differentiating the monomials 
of rank 1 in ${\cal M}$ with respect to $x$ twice, or $y$ twice, or once 
with respect to $x$ and $y.$ 
Again \mbox{ignore} the highest-order terms.
Augment the set ${\cal M}$ with the derivative terms of rank 3 to get
${\cal R} = \{ \Omega u, \Omega v, \cdots, u v^2, u_x v, u_x \theta, u_x h, 
\cdots, u_y v, u_y \theta, \cdots, \theta_y h \}$ 
which has 36 terms.

Instead of applying the algorithm of Section~\ref{algorithm} to ${\cal R},$
use the ``divide and conquer'' strategy to split ${\cal R}$ into sublists
of terms of equal rank under the (general) weights
\begin{eqnarray}
\label{swwscalegeneral}
&& W({\partial/\partial t}) =
W(\Omega), \quad 
W({\partial/\partial y}) = W({\partial/\partial x}) = 1, 
\quad W(u) = W(v) = W(\Omega) - 1, 
\nonumber \\
&& W(\theta) =
2 W(\Omega) - W(h) - 2, 
\end{eqnarray}  
where $W(\Omega)$ and $W(h)$ are arbitrary.
Use (\ref{swwscalegeneral}), to compute the rank of each monomial in 
${\cal R}$ and gather terms of like rank in separate lists.

Apply the algorithm from Section~\ref{algorithm} to each
${\cal R}_i$ to get the list ${\cal S}_i.$
Coincidentally, in this example ${\cal R}_i = {\cal S}_i$ for all $i.$ 
Linearly combine the monomials in each list ${\cal S}_i$ with coefficients 
to get the shortest candidate densities $\rho_i$.
In Table~\ref{swwcandidatedensities}, we list the 10 candidate densities and
the final densities with their ranks.
%
%
%
\vskip 0.0001pt
\noindent
\vspace{-0.750cm}
\begin{table}[h]
\begin{center}
\begin{tabular}{|l|l|l|l|l|}
\hline
\!i\!   & Rank 
    & Candidate ${\rho}_i$ 
    & Final ${\rho}_i$ 
    & Final ${\bf J}_i $
\\ \hline\hline
\!1\!   & $\!\!6W(\Omega) \!\!-\!\! 3 W(h) \!\!-\!\! 6\!\!$ 
    & $\!\!c_1 \theta^3 $ 
    & $\!\!0 $
    & $\!\!0 $
\\ \hline
\!2\!   & $\!\!3 W(h)$  
    & $\!\!c_1 h^3 $ 
    & $\!\!0 $
    & $\!\!0 $
\\ \hline
\!3\!   & $\!\!5 W(\Omega) \!\!-\!\! 2 W(h) \!\!-\!\!5\!\!$ 
    & $ \!\!c_1 u \theta^2 \!\!+\!\! c_2 v \theta^2 $
    & $\!\!0 $
    & $\!\!0 $
\\ \hline
\!4\!   & $\!\!W(\Omega) \!\!+\!\! 2 W(h) \!\!-\!\! 1$ 
    & $ \!\!c_1 u h^2 \!\!+\!\! c_2 v h^2 $
    & $\!\!0 $
    & $\!\!0 $
\\ \hline
\!5\!   & $\!\!4 W(\Omega) \!\!-\!\! W(h) \!\!-\!\! 4$ 
    & $ \!\!c_1 u^2 \theta \!\!+\!\! c_2 u v \theta 
        \!\!+\!\! c_3 v^2 \theta \!\!+\!\! c_4 \theta^2 h $
    & $\!\!\theta^2 h $
    & $\!\!\left(\!
        \begin{array}{c}
        u h \theta^2  \\
        v h \theta^2 
        \end{array}
        \!\right)\!\!$
\\ \hline
\!6\!   & $\!\!2 W(\Omega) \!\!+\!\! W(h) \!\!-\!\! 2$ 
    & $ \!\!c_1 u^2 h \!\!+\!\! c_2 u v h 
        \!\!+\!\! c_3 v^2 h \!\!+\!\! c_4 \theta h^2 $
    & $\!\!u^2 h \!\!+\!\! v^2 h \!\!+\!\! \theta h^2\!\!$
    & $\!\!\!\left(\!
        \begin{array}{c}
         u^3 h \!\!+\!\! u v^2 h \!\!+\!\! 2 u h^2 \theta \\
         v^3 h \!\!+\!\! u^2 v h \!\!+\!\! 2 v h^2 \theta 
         \end{array}
         \!\right)\!\!\!$
\\ \hline
\!7\!   & $\!\!3 W(\Omega) \!\!-\!\! W(h) \!\!-\!\! 2$ 
    & $ \!\!c_1 \Omega \theta \!\!+\!\! c_2 u_y \theta 
        \!\!+\!\! c_3 v_y \theta \!\!+\!\!c_4 u_x \theta 
        \!\!+\!\! c_5 v_x \theta$
    & $\!\!2 \Omega \theta  \!\!-\!\! u_y \theta \!\!+\!\! v_x \theta\!\!$
    & $\!\!{\bf J}_7$ in (\ref{J7sww})
\\ \hline
\!8\!   & $\!\!W(\Omega) \!\!+\!\! W(h)$ 
    & $ \!\!c_1 \Omega h \!\!+\!\! c_2 u_y h 
        \!\!+\!\! c_3 v_y h \!\!+\!\! c_4 u_x h \!\!+\!\! c_5 v_x h $
    & $\!\!\Omega h $
    & $\!\!\left(
        \begin{array}{c}
        \Omega u h  \\
        \Omega v h  
        \end{array}
        \right)\!\!$
\\ \hline
\!9\!   & $\!\!2 W(\Omega) \!\!-\!\! 1$  
    & $ \!\!c_1 \Omega u \!\!+\!\! c_2 \Omega v 
        \!\!+\!\! c_3 u_y v \!\!+\!\! c_4 \theta_y h
        \!\!+\!\! c_5 u_x v \!\!+\!\! c_6 \theta_x h\!\!$
    & $\!\!0 $
    & $\!\!0 $
\\ \hline
$\!\!10\!\!$& $\!\!3 W(\Omega) \!\!-\!\! 3$ 
    & $ \!\!c_1 u^3 \!\!+\!\! c_2 u^2 v 
        \!\!+\!\! c_3 u v^2 \!\!+\!\! c_4 v^3 \!\!+\!\! c_5 u \theta h
        \!\!+\!\! c_6 v \theta h\!\!$
    & $\!\!0 $
    & $\!\!0 $
\\ \hline
\end{tabular}
\caption{
\label{swwcandidatedensities}
Candidate densities for the SWW equations.
}
\vspace{-0.75cm}
\end{center}
\end{table}
%
%
\vskip 4pt
\noindent
{\bf Step 2}: {\bf Determine the constants $c_i$}
\vskip 2pt
\noindent
For each of the densities $\rho_i$ in Table~\ref{swwcandidatedensities}
compute $E_i \!=\! {\rm D}_t \rho_i$ and use (\ref{sww}) to remove all 
time derivatives.
For example, proceeding with $\rho_7,$ 
\begin{eqnarray}
\label{E7rho7sww}
E_7 &\!\!=\!\!&\! 
{\rho_7}^{\prime}({\bf u}) [{\bf F}] 
\!=\! 
\frac{\partial \rho_7}{\partial u_{x}} u_{tx}
+ \frac{\partial \rho_7}{\partial u_{y}} u_{ty}
+ \frac{\partial \rho_7}{\partial v_{x}} v_{tx}
+ \frac{\partial \rho_7}{\partial v_{y}} v_{ty}
+ \frac{\partial \rho_7}{\partial \theta } \theta_t 
\nonumber \\
&\!\!=\!\!&\! 
- c_4 \theta (u u_x + v u_y - 2 \Omega v + 
    {\textstyle \frac{1}{2}} h \theta_x + \theta h_x)_x
- c_2 \theta (u u_x + v u_y - 2 \Omega v + 
    {\textstyle \frac{1}{2}} h \theta_x + \theta h_x)_y
\nonumber \\
&\!\!\!\!&\! 
- c_5 \theta (u v_x + v v_y + 2 \Omega u + 
    {\textstyle \frac{1}{2}} h \theta_y + \theta h_y)_x
- c_3 \theta (u v_x + v v_y + 2 \Omega u + 
    {\textstyle \frac{1}{2}} h \theta_y + \theta h_y)_y 
\nonumber \\
&\!\!\!\!&\! 
- (c_1 \Omega + c_2 u_y + c_3 v_y + c_4 u_x + c_5 v_x)
  (u \theta_x + v \theta_y) .
\end{eqnarray}
Require that 
${\cal L}^{(0,0)}_{u(x,y)} (E_{7}) 
= {\cal L}^{(0,0)}_{v(x,y)} (E_{7}) 
= {\cal L}^{(0,0)}_{\theta(x,y)} (E_{7}) 
= {\cal L}^{(0,0)}_{h(x,y)} (E_{7}) \equiv 0,$
where, for example, ${\cal L}^{(0,0)}_{u(x,y)}$ is given in 
(\ref{zeroeulerscalaruxy}).
Gather like terms. 
Equate their coefficients to zero to obtain 
\begin{equation}
\label{systemrho7sww}
c_1 + 2 c_2 = 0, \quad c_3 = c_4 = 0, \quad
c_1 - 2 c_5 = 0, \quad c_2 + c_5 = 0.
\end{equation}
Set $c_1 = 2.$ 
Substitute the solution
\begin{equation}
\label{solrho7sww}
c_1 = 2, \; c_2 = -1, \; c_3 = c_4 = 0, \; c_5 = 1.
\end{equation}
into $\rho_7$ to obtain
$ \rho_7 = 2 \Omega \theta - u_y \theta + v_x \theta, $
which corresponds to $\rho^{(5)}$ in (\ref{swwconslaw}).

Proceed in a similar way with the remaining nine candidate densities 
to obtain the results given in the third column of 
Table~\ref{swwcandidatedensities}.
\vskip 4pt
\noindent
{\bf Step 3}: {\bf Compute the flux ${\bf J}$}
\vskip 2pt
\noindent
Compute the flux corresponding to all $\rho_i \ne 0$ in 
Table~\ref{swwcandidatedensities}.
For example, continuing with $\rho_7$, 
substitute (\ref{solrho7sww}) into (\ref{E7rho7sww}) to get
\begin{eqnarray}
\label{E7rho7swwupdate}
E_7 \!&\!=\!&\! 
%
%
%
- \theta (u_x v_x + u v_{2x} + v_x v_y + v v_{xy} + 2 \Omega u_x  
  +\! {\textstyle \frac{1}{2}} \theta_x h_y 
  - u_x u_y - u u_{xy} - u_y v_y - u_{2y} v
\nonumber \\
&\!\!\!\!&\! 
 + 2 \Omega v_y -\! {\textstyle \frac{1}{2}} \theta_y h_x ) 
- (2 \Omega u \theta_x + 2 \Omega v \theta_y - u u_y \theta_x 
  - u_y v \theta_y + u v_x \theta_x + v v_x \theta_y).
\end{eqnarray}
Apply the 2D homotopy operator in
(\ref{homotopyvectoruxycompx})-(\ref{homotopyvectoruxycompy})
to $E_7 = - {\rm Div} \, {\bf J}_7.$ 
So, compute 
\begin{eqnarray}
\nonumber
I_u^{(x)}(E_7) 
&=& u {\cal L}^{(1,0)}_{u(x,y)}(E_7)
+ {\rm D}_x \left( u {\cal L}^{(2,0)}_{u(x,y)}(E_7) \right)
+ \frac{1}{2}{\rm D}_y \left( u {\cal L}^{(1,1)}_{u(x,y)}(E_7) \right)
\\ \nonumber
&=& u \left( \frac{\partial E_7}{\partial u_x}
- 2 {\rm D}_x \left( \frac{\partial E_7}{\partial u_{2x}} \right)
- {\rm D}_y \left( \frac{\partial E_7}{\partial u_{xy}} \right) \right)
+ {\rm D}_x \left( u \frac{\partial E_7}{\partial u_{2x}} \right)
+ \frac{1}{2}{\rm D}_y \left( u \frac{\partial E_7}{\partial u_{xy}} \right)
\\
&=&
- u v_x \theta - 2 \Omega u \theta - \frac{1}{2} u^2 \theta_y + u u_y \theta.
\end{eqnarray}
Similarly, compute
\begin{eqnarray}
I_v^{(x)}(E_{7}) &=& 
 - v v_y \theta - \frac{1}{2} v^2 \theta_y - u v_x \theta,
\\
I_{\theta}^{(x)}(E_{7}) &=&
 -\frac{1}{2} \theta^2 h_y -2 \Omega u \theta + u u_y \theta - u v_x \theta, 
\\
I_{h}^{(x)}(E_{7}) &=&
 \frac{1}{2} \theta \theta_y h.
\end{eqnarray}
Next, compute 
\begin{eqnarray}
\label{dellarhomotop2Dxpart}
J_{7}^{(x)}({\bf u}) 
&\!\!=\!\!& -{\cal H}^{(x)}_{{\bf u}(x,y)} (E_{7}) 
\nonumber \\
& \!\!=\!\!&
- \int_{0}^{1} \left( 
  I_u^{(x)}(E_{7}) [\lambda {\bf u}] 
+ I_v^{(x)}(E_{7}) [\lambda {\bf u}] 
+ I_{\theta}^{(x)}(E_{7}) [\lambda {\bf u}] 
+ I_h^{(x)}(E_{7}) [\lambda {\bf u}] 
\right) \, \frac{d \lambda}{\lambda}
\nonumber \\
&\!\!=\!\!& \int_0^1\!
\left( 4 \lambda \Omega u \theta + 
\lambda^2 \left( 
3 u v_x \theta + \frac{1}{2} u^2 \theta_y - 2 u u_y \theta
+ v v_y \theta + \frac{1}{2} v^2 \theta_y 
+ \frac{1}{2} \theta^2 h_y - \frac{1}{2} \theta \theta_y h
\right) \right) \, d\lambda  
\nonumber \\ 
&\!\!=\!\!& 
2 \Omega u \theta - \frac{2}{3} u u_y \theta +  u v_x \theta 
+ \frac{1}{3} v v_y \theta  + \frac{1}{6} u^2 \theta_y 
+ \frac{1}{6} v^2 \theta_y - \frac{1}{6} h \theta \theta_y 
+ \frac{1}{6} h_y \theta^2.
\end{eqnarray}
In analogous fashion, compute
\begin{eqnarray}
\label{dellarhomotop2Dypart}
\!\!\!\!\! 
J_{7}^{(y)}({\bf u}) 
&\!\!=\!\!& -{\cal H}^{(y)}_{{\bf u}(x,y)} (E_{7}) 
\nonumber \\ 
&\!\!=\!\!& 
2 \Omega v \theta + \frac{2}{3} v v_x \theta - v u_y \theta 
- \frac{1}{3} u u_x \theta  - \frac{1}{6} u^2 \theta_x 
- \frac{1}{6} v^2 \theta_x + \frac{1}{6} h \theta \theta_x 
- \frac{1}{6} h_x \theta^2.
\end{eqnarray}
Hence, 
\begin{equation}
\label{J7sww}
{\bf J}_7 =
\frac{1}{6} \!\left(\!
 \begin{array}{c}
 12 \Omega u \theta - 4 u u_y \theta + 6 u v_x \theta + 2 v v_y \theta 
 + u^2 \theta_y + v^2 \theta_y - h \theta \theta_y + h_y \theta^2 
 \\
 12 \Omega v \theta + 4 v v_x \theta - 6 v u_y \theta - 2 u u_x \theta 
 - u^2 \theta_x - v^2 \theta_x + h \theta \theta_x - h_x \theta^2 
 \end{array}
 \!\right), 
\end{equation}
which matches ${\bf J}^{(5)}$ in (\ref{swwconslawtris}).

Proceed in a similar way with the remaining nonzero densities to obtain the 
fluxes given in the last column of Table~\ref{swwcandidatedensities}.

System (\ref{sww}) has conserved densities of the form
\begin{equation}
\label{swwgeneraldensities}
\rho = h f(\theta)
\quad {\rm and} 
\quad \rho = (v_x - u_y + 2 \Omega) g(\theta), 
\end{equation}
where $f$ and $g$ are arbitrary functions. 
Our algorithm can only find polynomial $f$ and $g.$
A comprehensive study of all conservation laws of (\ref{sww}) is 
beyond the scope of this chapter.
%
%
%
\section{Examples of Nonlinear DDEs}
\label{discreteexamples}
We consider nonlinear systems of DDEs of the form
\begin{equation}
\label{DDEsystem}
{\dot{\bf u}}_n ={\bf G} (...,{\bf u}_{n-1}, {\bf u}_{n}, {\bf u}_{n+1},...), 
\end{equation}
where ${\bf u}_{n}$ and ${\bf G}$ are vector-valued functions with $N$ 
components.
The integer $n$ corresponds to discretization in 
space\footnote{We only consider DDEs with one discrete variable.}; 
the dot denotes differentiation with respect to the continuous time $(t).$
For simplicity, we write ${\bf G}({\bf u}_n),$ although ${\bf G}$ 
depends on ${\bf u}_n$ and a finite number of its forward and backward shifts.
We assume that ${\bf G}$ is polynomial with constant coefficients.
No restrictions are imposed on the forward or backward shifts or the 
degree of nonlinearity in ${\bf G}.$
In the examples we denote the components of ${\bf u}_n$ by $u_n,v_n,$ etc. 
If present, parameters are denoted by lower-case Greek letters.
We use the following two DDEs to illustrate the theorems and algorithms.
\vskip 4pt
\noindent
{\bf Example 4:}
The {\rm Kac-van Moerbeke (KvM) lattice} \cite{MKandPvMam75},
\begin{equation}
\label{kvmlattice}
{\dot {u}}_n = u_n (u_{n+1} - u_{n-1}),
\end{equation}
arises in the study of Langmuir oscillations in plasmas, population 
dynamics, etc.\
\vskip 4pt
\noindent
{\bf Example 5:} The {\rm Toda lattice} \cite{MTbook81} in polynomial form 
\cite{UGandWHpd98},
\begin{equation}
\label{todalattice}
{\dot{u}}_n = v_{n-1} - v_n, \quad {\dot{v}}_n = v_n (u_n - u_{n+1}),
\end{equation}
models vibrations of masses in a lattice with an exponential 
interaction force.
%
\vspace{-0.50cm}
\section{Dilation Invariance and Uniformity in Rank for DDEs}
\label{discretedilation}
The definitions for the discrete case are analogous to the continuous case. 
For brevity, we use Example 5 to illustrate the definitions and concepts.
\vskip 1pt
\noindent
As shown in Table~\ref{examplesDDEs}, the Toda lattice (\ref{todalattice}) 
is invariant under the {\it scaling symmetry} 
\begin{equation}
\label{scaletoda}
(t, u_n, v_n) \rightarrow (\lambda^{-1} t,\lambda u_n,{\lambda}^2 v_n).
\end{equation}
\vskip 1pt
\noindent
{\bf Definition}: 
The {\it weight} $W$ of a variable equals the exponent of the scaling 
parameter $\lambda.$ 
\cite{UGandWHpd98,UGandWHacm99}.
Weights of dependent variables are nonnegative and rational.
We tacitly assume that weights are independent of $n$. 
For example, $W(u_{n-1}) = W(u_n) = W(u_{n+1}),$ etc.
\vskip 3pt
\noindent
{\bf Example}: 
Since $t$ is replaced by 
$\frac{t}{\lambda}$ we have 
$W(\frac{\rm{d}}{\rm{d}t}) = W({\rm D}_t) = 1.$ 
From (\ref{scaletoda}) we have $W(u_n)=1$ and $W(v_n)=2.$
\vskip 3pt
\noindent
{\bf Definition}: 
The {\it rank} of a monomial equals the total weight of the monomial. 
An expression (or equation) is {\rm uniform in rank} if all its monomial 
terms have equal rank.
\vskip 3pt
\noindent
{\bf Example}: 
The three terms in the first equation in (\ref{todalattice}) have rank 2; 
all terms in the second equation have rank 3.
Each equation is uniform in rank.

Conversely, requiring uniformity in rank in (\ref{todalattice}) yields
$W(u_n)+1\!=\!W(v_n),$ and $W(v_n)+1\!=\!W(u_n)+W(v_n).$
Hence, $W(u_n)\!=\!1,\,W(v_n)\!=\!2.$ 
So, the scaling symmetry can be computed with linear algebra. 

Many integrable nonlinear DDEs are scaling invariant.
If not, they can be made so by extending the set of dependent variables 
with parameters with weights.
\vspace{-0.50cm}
\section{Conserved Densities and Fluxes of Nonlinear DDEs}
\label{densfluxDDEs}
%
By analogy with ${\rm D}_x$ and ${\rm D}_x^{-1},$
we define the following operators acting on monomials $m_n$ in $u_n,v_n,$ etc.
\vskip 3pt
\noindent
{\bf Definition}: 
${\rm D}$ is the {\it up-shift operator} 
(also known as the forward- or right-shift operator) 
${\rm D} \, m_n = m_{n+1}.$ 
Its inverse, ${\rm D}^{-1},$ is the {\it down-shift operator} 
(or backward- or left-shift operator), ${\rm D^{-1}} \, m_n = m_{n-1}.$ 
The {\rm identity operator} is denoted by ${\rm I},$ ${\rm I} \, m_n = m_n$
and $\Delta = {\rm D} - {\rm I},$ is the {\rm forward difference operator}.
So, $\Delta \, m_n = ({\rm D} - {\rm I}) \, m_n = m_{n+1} - m_n.$
\vskip 3pt
\noindent
{\bf Definition}: 
A {\it conservation law} of (\ref{DDEsystem}),
\begin{equation}
\label{ddeconslaw}
{\rm D}_t \, \rho_n + \Delta \, J_n = 0,
\end{equation}
which holds on solutions of (\ref{DDEsystem}), 
links a {\it conserved density} $\rho_n$ to a {\it flux} $J_n.$
Densities and fluxes depend on ${\bf u}_n$ as well as forward and 
backward shifts of ${\bf u}_n.$
\vskip 3pt
\noindent
{\bf Definition}: 
Compositions of ${\rm D}$ and ${\rm D}^{-1}$ define an 
{\it equivalence relation} $(\equiv)$ on monomials.
All shifted monomials are equivalent.
\vskip 3pt
\noindent
{\bf Example}: 
For example, 
$u_{n-1} v_{n+1} \equiv u_{n} v_{n+2} \equiv u_{n+1} v_{n+3} 
\equiv u_{n+2} v_{n+4}.$
Factors in a monomial in $u_n$ and its shifts are ordered by
$u_{n+j} \prec u_{n+k}$ if $j < k.$
\vskip 3pt
\noindent
{\bf Definition}: 
The {\it main representative} of an equivalence class 
is the monomial with $u_n$ in the first position 
\cite{UGandWHpd98,UGandWHacm99}. 
\vskip 3pt
\noindent
{\bf Example}: 
The main representative in class
$\{\cdots, u_{n-2} u_{n}, u_{n-1} u_{n+1},$ $u_n u_{n+2}, \cdots\}$
is $u_n u_{n+2}$ (not $u_{n-2} u_n).$  

For monomials involving $u_n, v_n, w_n,$ etc.\ and their shifts, 
we lexicographically order the variables, i.e.\ 
$u_n \prec v_n \prec w_n,$ etc.\ 
For example, $u_n v_{n+2} $ (not $u_{n-2} v_n$) is the main representative of
$\{\cdots, 
u_{n-2} v_{n}, u_{n-1} v_{n+1}, u_{n} v_{n+2}, u_{n+1} v_{n+3}, 
\cdots \}.$ 
\vskip 3pt
\noindent
%
To stress the analogy between PDEs and DDEs, we put the defining equations 
next to each other in Table~\ref{analogypdedde}.
%
%
\vspace{-0.450cm}
\vskip 1pt
\noindent
\begin{center}
\begin{table}[h]
\begin{center}
\begin{tabular}{|l|l|l|} \hline 
& Continuous $\!$Case $\!$(PDE) & Semi-discrete Case (DDE) \\ [0.6ex] 
\hline
$\!\!$Evolution Equation & 
${\bf u}_t={\bf G}({\bf u}, {\bf u}_{x},{\bf u}_{y},...,{\bf u}_{2x},...)$ &
${\dot{\bf u}}_n\!\!=\!\!{\bf G}(...,
{\bf u}_{n-1}, {\bf u}_{n}, {\bf u}_{n+1},...)$ \\ [0.6ex] 
\hline 
$\!\!$Conservation Law & $ {\rm D}_{t} \rho + 
{\mbox {\boldmath $\nabla$}} \cdot {\bf J} = 0 $ &
${\dot{\rho}}_n + \Delta \, J_n = 0 $ \\ [0.6ex]
\hline 
\end{tabular}
\caption{
\label{analogypdedde}
Defining equations for conservation laws for PDEs and DDEs.}
\end{center}
\end{table}
\end{center}
\vspace{-0.550cm}
%
%
%
Table~\ref{examplesDDEs} shows the KvM and Toda lattices with their 
scaling invariance (and weights) and a few conserved densities.
Notice that the conservation law ``inherits" the scaling symmetry of the DDE.
Indeed, observe that all $\rho_n$ in Table~\ref{examplesDDEs} are uniform 
in rank, be it of different ranks.
%
\vspace{-0.650cm}
\begin{table}[h]
\begin{center}
\begin{tabular}{|l|l|l|} \hline 
& Kac-van Moerbeke Lattice & Toda Lattice 
\\ [0.5ex] \hline
$\!$Lattice$\!$& ${\dot{u}}_n = u_n (u_{n+1}-u_{n-1})$ & 
${\dot{u}}_n =v_{n-1}-v_n,\quad {\dot{v}}_n = v_n (u_n - u_{n+1})$ 
\\ [0.5ex] \hline
$\!$Scaling$\!$ 
& $(t, u_n) \rightarrow (\lambda^{-1} t, \lambda u_n)$ 
& $(t,u_n,v_n) \rightarrow (\lambda^{-1} t, \lambda u_n, {\lambda}^2 v_n)$ 
\\ [0.5ex] \hline 
$\!$Weights$\!$
& $W({\rm D}_t)=1, \, W(u_n) = 1$
& $W({\rm D}_t)=1, \, W(u_n)= 1, \, W(v_n) = 2$ 
\\ [0.5ex] \hline
$\!$Densities & 
  ${\rho}_n^{(1)} = u_n, \quad 
   {\rho}_n^{(2)} = \frac{1}{2} u_n^2 \!+ u_n u_{n+1} $
& ${\rho}_n^{(1)} = u_n, \quad 
   {\rho}_n^{(2)} = \frac{1}{2} u_n^2 + v_n $ 
\\ [0.65ex]  
   & ${\rho}_n^{(3)} \!=\! \frac{1}{3} u_n^3 \!+\! u_n u_{n+1} 
     (u_n + u_{n+1} \!+\! u_{n+2})\!$
   & ${\rho}_n^{(3)} =\frac{1}{3} u_n^3 + u_n ( v_{n-1} + v_n ) $ 
\\ [0.5ex]
\hline 
\end{tabular}
\caption{
\label{examplesDDEs}
Examples of nonlinear DDEs with weights and densities.}
\end{center}
\end{table}
\vspace{-0.950cm}
%
\section{Discrete Euler and Homotopy Operators}
\label{toolsdiscrete}
\subsection{\bf Discrete Variational Derivative (Euler Operator)}
\label{eulerdiscrete}
Given is a scalar function $f_n$ in discrete variables $u_n, v_n,\ldots $ 
and their forward and backward shifts.
The goal is to find the scalar function $F_n$ so that 
$f_n = {\rm \Delta} \, F_n = F_{n+1} - F_n.$ 
We illustrate the computations with the following example: 
\begin{equation}
\label{fdiscrete}
f_n \!=\!
   - u_n u_{n+1}  v_n - v_n^2 + u_{n+1}  u_{n+2} v_{n+1} 
   + v_{n+1}^2 + u_{n+3} v_{n+2} - u_{n+1} v_n.
\end{equation}
By hand, one readily computes
\begin{equation}
\label{resultbyhand}
F_n = v_n^2 +  u_n \, u_{n+1} \, v_n +  u_{n+1} \, v_n + u_{n+2} \, v_{n+1}.
\end{equation}
Below we will address the questions: 
\vskip 1pt
\noindent
(i) Under what conditions for $f_n$ does $F_n$ exist in closed form?
\vskip 1pt
\noindent
(ii) How can one compute $F_n = {\rm \Delta}^{-1}(f_n)\, ?$
\vskip 1pt
\noindent
(iii) Can one compute $F_n = \Delta^{-1}(f_n)$ in an analogous way as in 
the continuous case? 
%
%

Expression $f_n$ is called {\it exact} if it is a total difference, 
i.e.\ there exists a $F_n$ so that $f_n = {\rm \Delta}\, F_n.$
With respect to the existence of $F_n$ in closed form, the following 
exactness criterion is well-known and frequently used 
\cite{VAetaltmp00,MHandWHprsa03}. 
\vskip 2pt
\noindent
{\bf Theorem}: 
A necessary and sufficient condition for a function $f_n,$ with positive 
shifts, 
to be exact is that ${\cal L}^{(0)}_{{\bf u}_n} (f_n) \equiv 0.$
\vskip 4pt
\noindent
${\cal L}^{(0)}_{{\bf u}_n}$ is the {\it variational derivative}
(discrete Euler operator of order zero) \cite{VAetaltmp00} defined by
\begin{equation}
\label{discreteeuleroperator}
{\mathcal L}^{(0)}_{{\bf u}_n} =
\frac{\partial }{\partial {\bf u}_{n}} ( \sum_{k=0}^{\infty} {\rm D}^{-k} )
= \frac{\partial }{\partial{{\bf u}_{n}}} 
( {\rm I} + {\rm D}^{-1} + {\rm D}^{-2} + {\rm D}^{-3} + \cdots ).
\end{equation}
A proof of the theorem is given in e.g.\ \cite{MHandWHprsa03}.
In practice, the series in (\ref{discreteeuleroperator}) terminates at the 
highest shift occurring in the expression the operator is applied to.
To verify that an expression $E(u_{n-q}, \cdots, u_n, \cdots, u_{n+p})$ 
involving negative shifts is a total difference, one must first remove the 
negative shifts by replacing $E_n$ by ${\tilde E}_n = {\rm D}^q E_n.$
\vskip 3pt
\noindent
{\bf Example:} 
We return to (\ref{fdiscrete}), 
$f_n = - u_n \, u_{n+1} \, v_n - v_n^2 + u_{n+1} \, u_{n+2} \, v_{n+1} 
+ v_{n+1}^2 + u_{n+3} \, v_{n+2} - u_{n+1} \, v_n. $
We first test if $f_n$ is exact (i.e.,\ the total difference of some $F_n$ 
to be computed later). 
We apply the discrete zeroth Euler operator to $f_n$ for each
component of ${\bf u}_n = (u_n,v_n)$ separately.
For component $u_n$ (with maximum shift 3)
one readily verifies that 
\begin{equation}
\label{eulerzeroforun}
{\cal L}^{(0)}_{{u}_n} (f_n) = \frac{\partial }{\partial{u_n}} 
\left( {\rm I} + {\rm D}^{-1} + {\rm D}^{-2} + {\rm D}^{-3} \right) (f_n) 
\equiv 0.
\end{equation}
Similarly, for component $v_n$ (with maximum shift 2)
one checks that
${\cal L}^{(0)}_{{v}_n} (f_n) \equiv 0.$  
%
%
%
%
%
%
\subsection{Discrete Higher Euler and Homotopy Operators}
\label{highereulerhomotopydiscrete}
%
To compute $F_n,$ we need higher-order versions of the discrete 
variational derivative. 
They are called {\it discrete higher Euler operators} 
${\cal L}^{(i)}_{{\bf u}_n}$ 
in analogy with the continuous case \cite{PObook93}. 
\vskip 1pt
\noindent
In Table~\ref{eulerandhomotopyoperators}, we have put the 
continuous and discrete higher Euler operators side by side.
Note that the discrete higher Euler operator for $i=0$ is the 
discrete variational derivative.
%
%
%
%
%
%
The first three higher Euler operators for component $u_n$ from 
Table~\ref{eulerandhomotopyoperators} are
\begin{eqnarray}
{\cal L}^{(1)}_{u_n} 
\!\!&\!=\!&\!\!
\frac{\partial }{\partial {u_n}} 
\left( {\rm D}^{-1} + 2 {\rm D}^{-2} + 3 {\rm D}^{-3} + 
4 {\rm D}^{-4} + \cdots \right), 
\\
{\cal L}^{(2)}_{u_n} 
\!\!&\!=\!&\!\!
\frac{\partial }{\partial{u_n}} 
\left( {\rm D}^{-2} + 3 {\rm D}^{-3} + 6 {\rm D}^{-4} + 
10 {\rm D}^{-5} + \cdots \right). 
\\
{\cal L}^{(3)}_{u_n} 
\!\!&\!=\!&\!\!
\frac{\partial }{\partial{u_n}} 
\left( {\rm D}^{-3} + 4 {\rm D}^{-4} + 10 {\rm D}^{-5} + 
20 {\rm D}^{-6} + \cdots \right) . 
\end{eqnarray}
Similar formulae hold for ${\mathcal L}^{(i)}_{v_n}.$
%
\vspace{-0.950cm}
\noindent
\begin{table}[b]
\begin{center}
\begin{tabular}{|l|l|l|} \hline 
& $\!$ Continuous Case $\!$ & $\!$Discrete Case$\!$ 
\\ [0.5ex] \hline
$\!$Zeroth Euler Operator$\!$
 & ${\cal L}^{(0)}_{{\bf u}(x)} \!\!=\!\! \sum_{k=0}^{\infty} 
   (-{\rm D}_x)^k \frac{\partial }{\partial {\bf u}^{(k)}} $
 & ${\mathcal L}^{(0)}_{{\bf u}_n} \!\!=\!\! \sum_{k=0}^{\infty} 
   {{\rm D}}^{-k} \frac{\partial }{\partial {\bf u}_{n+k}} $ 
\\ [0.7ex] \hline
$\!$Higher Euler Operators $\!$
 & ${\cal L}^{(i)}_{{\bf u}(x)} \!\!=\!\! \sum_{k=i}^{\infty} 
   {k \choose i} (-{\rm D}_x)^{k-i} \frac{\partial }{\partial {\bf u}^{(k)}}$
 & ${\cal L}^{(i)}_{{\bf u}_n} \!\!=\!\! \frac{\partial }{\partial{{\bf u}_n}} 
\left( \sum_{k=i}^{\infty} {k \choose i} {\rm D}^{-k} \right)$
\\ [0.7ex] \hline
$\!$Homotopy Operators $\!$
 & $ {\cal H}_{{\bf u}(x)} (f) 
\!\!=\!\! \int_{0}^{1} \sum_{j=1}^{N} I_{u_j} (f)
 [\lambda {\bf u}] \frac{d \lambda}{\lambda}$
 & ${\cal H}_{{\bf u}_n} (f) \!\!=\!\! \int_{0}^{1} \sum_{j=1}^{N} 
   I_{u_{j,n}} (f)[\lambda {\bf u}_n] \frac{d \lambda}{\lambda}$ 
\\ [0.7ex] \hline
$\!$Integrand Operator $\!$
 & $ I_{u_j} (f) \!\!=\!\! \sum_{i=0}^{\infty} {\rm D}_{x}^i 
    \left( u_j {\cal L}^{(i+1)}_{{u_j}(x)} (f) \right) $
 & $I_{u_{j,n}}(f) \!\!=\!\! \sum_{i=0}^{\infty} \Delta^i 
    \left( u_{j,n} {\cal L}^{(i+1)}_{u_{j,n}} (f) \right) $
\\ [0.7ex] \hline
\end{tabular}
\caption{
\label{eulerandhomotopyoperators}
Continuous and discrete Euler and homotopy operators in 1D side by side.}
\end{center}
\end{table}
\vspace{-1.00cm}
%
\newpage
\newpage
\noindent
Also in Table~\ref{eulerandhomotopyoperators}, we put the formulae for the 
{\it discrete homotopy operator} ${\cal H}_{{\bf u}_n}$ 
and the continuous homotopy operator side by side.
The integrand $I_{u_{j,n}} (f)$ of the homotopy operator involves the 
discrete higher Euler operators. 
%
%
As is the continuous case, $N$ is the number of dependent variables and 
$I_{u_{j,n}} (f)[\lambda {\bf u}_n]$ means that after 
$I_{u_{j,n}} (f)$ is applied one replaces ${\bf u}_n$ 
by $\lambda {\bf u}_n,\, {\bf u}_{n+1}$ by $ \lambda {\bf u}_{n+1},$ etc.
%
%
%
%
%
\noindent
To compute $F_n,$ one can use the following theorem 
\cite{WHetalcrm04,PHandEMfcm04,EMandPHams02}.
\vskip 2pt
\noindent
\noindent
{\bf Theorem}: 
Given an exact function $f_n,$ one can compute $F_n = \Delta^{-1}(f_n)$ 
from $F_n = {\mathcal H}_{{\bf u}_n} (f_n).$
%
\vskip 4pt
\noindent
Thus, the homotopy operator reduces
the inversion of $\Delta$ (and summation by parts) to a set of 
differentiations and shifts followed by a single integral with respect 
to an auxiliary parameter $\lambda.$
We present a simplified version \cite{WHetalcrm04} of the homotopy operator 
given in \cite{PHandEMfcm04,EMandPHams02}, where the problem is dealt with 
in greater generality and where the proofs are given in context of 
variational complexes.
%
%
%
%

For a system with components, $(u_{1,n}, u_{2,n})=(u_n,v_n),$ 
the discrete homotopy operator from Table~\ref{eulerandhomotopyoperators} is
\begin{equation}
\label{discretetotalhomotopyoperator2}
{\cal H}_{{\bf u}_n} (f) = \int_{0}^{1} 
\left( 
I_{u_n}(f)[\lambda {\bf u}_n] 
+  I_{v_n}(f)[\lambda {\bf u}_n] 
\right) \frac{d \lambda}{\lambda},
\end{equation}
with
\begin{equation}
\label{discretehomotopyoperator2}
I_{u_n}(f) = \sum_{i=0}^{\infty} 
\Delta^i \left( u_{n} {\cal L}^{(i+1)}_{u_{n}} (f) \right)
\quad\quad {\rm and} \quad\quad
I_{v_n}(f) = \sum_{i=0}^{\infty} 
\Delta^i \left( v_{n} {\cal L}^{(i+1)}_{v_{n}} (f) \right).
\end{equation}
\vskip 3pt
\noindent
{\bf Example:} 
We return to (\ref{fdiscrete}). 
Using (\ref{discretehomotopyoperator2}),
\begin{eqnarray}
\nonumber
I_{u_n}(f_n) &=&
u_n {\cal L}^{(1)}_{u_n}(f_n)+\Delta
\left( u_n {\cal L}^{(2)}_{u_n}(f_n) \right)
+ \Delta^2 \left( u_n {\cal L}^{(3)}_{u_n}(f_n) \right)
\\ \nonumber
&=& u_n \frac{\partial }{\partial u_n} \left({\rm D}^{-1} 
+ 2{\rm D}^{-2}+3{\rm D}^{-3} \right)(f_n) 
+ \Delta \left( u_n \frac{\partial }{\partial u_n}
\left( {\rm D}^{-2}+3{\rm D}^{-3}\right)(f_n) \right) 
\\ \nonumber
&& + \Delta^2\left(u_n\frac{\partial }{\partial u_n}{\rm D}^{-3}(f_n)\right)
\\
&=& 2 u_{n} u_{n+1} v_{n} + u_{n+1} v_{n} + u_{n+2} v_{n+1},
\end{eqnarray}
and
\begin{eqnarray}
\nonumber
I_{v_n}(f_n) &=& v_n {\cal L}^{(1)}_{v_n}(f_n)
+ \Delta \left( v_n {\cal L}^{(2)}_{v_n}(f_n) \right)
\\ \nonumber
&=& v_n \frac{\partial }{\partial v_n}\left({\rm D}^{-1} 
+ 2{\rm D}^{-2}\right)(f_n) 
+ \Delta \left( v_n \frac{\partial }{\partial v_n}{\rm D}^{-2}(f_n) \right)
\\
&=& u_{n} u_{n+1} v_{n} + 2 v_{n}^2 + u_{n+1} v_{n} + u_{n+2} v_{n+1}.
\end{eqnarray}
The homotopy operator (\ref{discretetotalhomotopyoperator2}) 
thus leads to an integral with respect to $\lambda:$
\begin{eqnarray}
F_n &=& 
\int_0^1 \left( I_{u_n}(f_n)[\lambda {\bf u}_n] 
+ I_{v_n}(f_n)[\lambda {\bf u}_n] \right) 
\; \frac{d\lambda}{\lambda} 
\nonumber \\
&=& \int_0^1 
\left( 
2 \lambda v_{n}^2 + 3 \lambda^2 u_{n} u_{n+1} v_{n} + 2 \lambda u_{n+1} v_{n} 
+ 2 \lambda u_{n+2} v_{n+1} \right) \; d\lambda  \\
&=& 
v_n^2 + u_{n} u_{n+1} v_{n} + u_{n+1} v_{n} + u_{n+2} v_{n+1},
\end{eqnarray}
which agrees with (\ref{resultbyhand}), previously computed by hand.
%
\section{Application: 
Conservation Laws of Nonlinear DDEs}
\label{applicationsDDEs}
In \cite{HEthesis03,MHandWHprsa03}, different algorithms are presented to 
compute fluxes of nonlinear DDEs.
In this section we show how to compute fluxes with the discrete 
homotopy operator.
%
%
%
%
For clarity, we compute a conservation law for Example 5 in
Section~\ref{discreteexamples}.
The computations are carried out with our {\it Mathematica} packages 
\cite{WHwebsite04}.
%
The completely integrable Toda lattice (\ref{todalattice}) has 
infinitely many conserved densities and fluxes. 
As an example, we compute density ${\rho}_n^{(3)}$ (of rank 3) 
and corresponding flux $J_n^{(3)}$ (of rank 4).
In this example, 
${\bf G} = (G_1, G_2) = (\, v_{n-1}-v_n, \, v_n (u_n - u_{n+1})\, ).$
%
%
%
Assuming that the weights $W(u_n)\!=\!1$ and $W(v_n)\!=\!2$ 
are computed and the rank of the density is selected (say,$R\!=\!3)$, 
our algorithm works as follows: 
%
\vskip 4pt
\noindent
{\bf Step 1: Construct the form of the density}
\vskip 2pt
\noindent
Start from ${\cal V} = \{ u_n, v_n \},$ i.e.\ the list of dependent 
variables with weight.
List all monomials in $u_n$ and $v_n$ of rank $3$ or less:
${\mathcal M} \!=\! \{ {u_n}^3, {u_n}^2, u_n v_n, {u_n}, v_n \}.$

Next, for each monomial in ${\mathcal M}$, introduce the correct number 
of $t$-derivatives so that each term has rank $3.$ 
Using (\ref{todalattice}), compute
\begin{eqnarray}
\label{todaweightadjust}
&&
\frac{{\rm d}^0{u_n}^3}{{\rm d}t^0} = {u_n}^3 , 
\;\;\;\;\quad 
\frac{{\rm d}^0 u_n v_n}{{\rm d}t^0} = u_n v_n , 
\nonumber \\
&&
\frac{{\rm d}{u_n}^2}{{\rm d}t} = 
2 u_n {\dot{u}}_n = 2 u_n v_{n-1} - 2 u_n v_n ,  
\;\;\;\;\quad
\frac{{\rm d} v_n}{{\rm d}t} = 
{\dot{v}}_n =  u_n v_n -  u_{n+1} v_n, 
\nonumber \\
&&
\frac{{\rm d}^2 u_n}{{\rm d}t^2} = \frac{{\rm d}{\dot{u}}_n}{{\rm d}t}  
   = \frac{{\rm d} (v_{n-1} - v_n)}{{\rm d}t} 
    = u_{n-1} v_{n-1} - u_{n} v_{n-1} - u_n v_n + u_{n+1} v_n .
\end{eqnarray}
Augment ${\mathcal M}$ with the terms from the right hand sides of 
(\ref{todaweightadjust}) to get 
\newline
$\mathcal R=$ 
$\{ {u_n}^3, u_n v_{n-1} , u_n v_n , u_{n-1} v_{n-1} , u_{n+1} v_n \}.$

Identify members belonging to the same equivalence classes and 
replace them by their main representatives.
For example, $u_n v_{n-1} \equiv u_{n+1} v_n,$ so the latter is replaced 
by $u_n v_{n-1}.$ 
Hence, replace $\mathcal R$ by 
${\mathcal S} = \{ {u_n}^3 , u_n v_{n-1} , u_n v_n \}, $
which has the building blocks of the density. 
Linearly combine the monomials in ${\mathcal S}$ with coefficients 
$c_i$ to get the candidate density:
\begin{equation}
\label{formrho3toda}
\rho_n = c_1 \, {u_n}^3 + c_2 \, u_n v_{n-1} + c_3 \, u_n v_n . 
\end{equation}
\vskip 4pt
\noindent
{\bf Step 2: Determine the coefficients}
\vskip 2pt
\noindent
Require that (\ref{ddeconslaw}) holds.
Compute ${\rm D}_t \rho_n.$ 
Use (\ref{todalattice}) to remove ${\dot{u}_n}$ and ${\dot{v}_n}$ 
and their shifts.
Thus, 
\begin{eqnarray}
\label{todalatticerhodot}
E_n \!&=&\! {\rm D}_t \rho_n =
( 3 c_1 - c_2 ) u_n^2 v_{n-1} + (c_3 - 3 c_1 ) u_n^2 v_n 
+ (c_3 - c_2) v_{n-1} v_n  
\nonumber \\
&& +\, c_2\, u_{n-1} u_n v_{n-1} + c_2 v_{n-1}^2 -c_3 u_n u_{n+1} v_n 
- c_3 v_n^2.
\end{eqnarray}
To remove the negative shift $n-1,$ compute ${\tilde E}_n = {\rm D} E_n$ 
Apply ${\cal L}^{(0)}_{u_n}$ to ${\tilde E}_n,$ yielding
\begin{eqnarray}
\label{applyeulerforun}
{\mathcal L}^{(0)}_{u_n} ({\tilde E}_n) 
\!&\!=\!&\!
\frac{\partial }{\partial {u_{n}}}
( {\rm I} + {\rm D}^{-1} + {\rm D}^{-2} ) ({\tilde E}_n) 
\nonumber \\
\!&\!=\!&\!
2 (3 c_1 - c_2 ) u_n v_{n-1} + 2 (c_3 - 3 c_1 ) u_n v_n 
+ ( c_2 - c_3 ) u_{n-1} v_{n-1}
\nonumber \\
&& + \, (c_2 - c_3) u_{n+1} v_{n}.
\end{eqnarray}
Next, apply ${\cal L}^{(0)}_{v_n}$ to ${\tilde E}_n,$ yielding
%
%
\begin{eqnarray}
\label{applyeulerforvn}
{\mathcal L}^{(0)}_{v_n} ({\tilde E}_n) 
\!&\!=\!&\!
\frac{\partial }{\partial {v_{n}}}( {\rm I} + {\rm D}^{-1} ) ({\tilde E}_n ) 
\nonumber \\
\!&\!=\!&\!
(3 c_1 - c_2 ) u_{n+1}^2 + (c_3 - c_2) v_{n+1} 
+ ( c_2 - c_3 ) u_{n} u_{n+1} + 2 (c_2 - c_3) v_{n}
\nonumber \\
&& 
+\, (c_3 - 3 c_1 ) u_n^2 + ( c_3 - c_2 ) v_{n-1}. 
\end{eqnarray}
Both (\ref{applyeulerforun}) and (\ref{applyeulerforvn}) must vanish 
identically. 
Solve the linear system
\begin{equation}
\label{todacsystem}
 3 c_1 - c_2 = 0, \quad c_3 - 3 c_1 = 0, \quad c_2 - c_3 = 0. 
\end{equation}
Set $c_1 = \frac{1}{3}$ and substitute the solution 
$c_1 = \frac{1}{3}, c_2 = c_3 = 1,$ into (\ref{formrho3toda})
\begin{equation}
\label{todacondens3final}
\rho_n = \frac{1}{3} {u_n}^3 + u_n ( v_{n-1} + v_n ).
\end{equation}
\vskip 4pt
\noindent
{\bf Step 3}: {\bf Compute the flux}
\vskip 2pt
\noindent
In view of (\ref{ddeconslaw}), one must compute $J_n = -\Delta^{-1}(E_n).$ 
%
%
Substitute $c_1 = \frac{1}{3}, c_2 = c_3 =1 $ into (\ref{todalatticerhodot}).
Then, ${\tilde E}_n = {\rm D} E_n = 
u_{n} u_{n+1} v_{n} + v_{n}^2 - u_{n+1} u_{n+2} v_{n+1} - v_{n+1}^2. $ 
\vskip 0.1pt
\noindent
Apply (\ref{discretehomotopyoperator2}) to $-{\tilde E}_n$ to obtain 
%
%
%
\begin{equation}
\label{todaf1andf2}
I_{u_n}(-{\tilde E}_n) = 2 u_n u_{n+1} v_n, \quad\;\;
I_{v_n}(-{\tilde E}_n) = u_n u_{n+1} v_n + 2 v_n^2.
\end{equation}
Application of homotopy operator (\ref{discretetotalhomotopyoperator2}) yields
\begin{eqnarray}
\label{todatotalhomotopy}
\tilde{J}_n &=& 
 \int_0^1 ( I_{u_n} (-{\tilde E}_n)[\lambda {\bf u}_n] + 
            I_{v_n} (-{\tilde E}_n)[\lambda {\bf u}_n]) \; 
\frac{d\lambda}{\lambda} 
\nonumber \\ 
&=&  \int_0^1 ( 3 \lambda^2 u_n u_{n+1} v_n + 2 \lambda v_n^2 ) \; d\lambda  
\nonumber \\
&=& u_n u_{n+1} v_n + v_n^2.
\end{eqnarray}
After a backward shift, $J_n = {\rm D}^{-1}({\tilde J}_n),$ we obtain $J_n.$
With (\ref{todacondens3final}), the final result is then
\begin{equation} 
\label{finalrhoandj}
\rho_n = \frac{1}{3} \, u_n^3 + u_n ( v_{n-1} + v_n ),
\quad 
J_n = u_{n-1} u_n v_{n-1} + v_{n-1}^2.
\end{equation}
The above density corresponds to $\rho_n^{(3)}$ in Table~\ref{examplesDDEs}.
%
\section{Conclusion}
\label{conclusion}
%
%
Based on the concept of scaling invariance and using tools of the
calculus of variations, we presented algorithms to symbolically compute 
conservation laws of nonlinear polynomial and transcendental systems of 
PDEs in multi-spacial dimensions and DDEs in one discrete variable.
We covered the symbolic computation of densities and fluxes. 

The continuous homotopy operator turned out to be a powerful, algorithmic
tool to compute fluxes explicitly. 
Indeed, the homotopy operator handles integration by parts in 
multi-variables which allowed us to invert the total divergence operator.
Likewise, the discrete homotopy operator handles summation by parts and 
inverts the forward difference operator. 
In both cases, the problem is reduced to an explicit integral from 1D calculus.

Homotopy operators have a wide range of applications in the study of 
PDEs, DDEs, fully discretized lattices, and beyond.
We extracted the continuous and discrete Euler and homotopy operators 
from pure mathematics, introduced them into applied mathematics, and 
therefore make them readily applicable to computational problems. 
We purposely avoided differential forms \cite{PObook93} and abstract concepts 
from differential geometry and homological algebra.

Our down-to-earth approach might appeal to scientists who prefer not to 
juggle exterior products and Lie derivatives.
%
Our calculus-based formulas for the Euler and homotopy operators can be 
readily implemented in major CAS.
%
%
\section*{Acknowledgements and Dedication}
\label{acknowledgements}
This material is based upon work supported by the National Science 
Foundation (NSF) under Grants Nos.\ DMS-9732069, DMS-9912293, 
CCR-9901929, and FRG-DMS-0139093. 
MN wishes to acknowledge support from NSF Grant No.\ DMS-9810726 (VIGRE).
Any opinions, findings, and conclusions or recommendations expressed in this
material are those of the authors and do not necessarily reflect the 
views of NSF. 

WH is grateful for the hospitality and support of the Department of 
Mathematics and Statistics of the University of Canterbury 
(Christchurch, New Zealand) where he was a Visiting Erskine Fellow in 
Summer 2004. 
The Erskine Foundation is thanked for financial support.

The authors are grateful to undergraduate students Lindsay Auble, 
Robert `Scott' Danford, Ingo Kabirschke, Forrest Lundstrom, Frances Martin, 
Kara Namanny, and Maxine von Eye for designing 
{\em Mathematica} code for the project.  


The research was supported in part by an Undergraduate Research Fund 
Award from the Center for Engineering Education awarded to RS.
On June 16, 2003, while rock climbing in Wyoming, Ryan was killed by a 
lightening strike. 
He was 20 years old. 
The authors express their gratitude for the insight Ryan brought to 
the project.
His creativity and passion for mathematics were inspiring.
This chapter honors his contribution to the project. 
%
%

\end{document}